\documentclass[journal]{IEEEtran}

\usepackage[english]{babel}
\usepackage[latin1]{inputenc}
\usepackage{enumerate}
\usepackage[T1]{fontenc}
\usepackage[T1]{fontenc}
\usepackage{amsmath}
\usepackage{amsthm}
\usepackage{amstext}
\usepackage{amssymb}
\usepackage{mathrsfs}
\usepackage{mathtools}
\usepackage{graphicx}
\usepackage{xcolor}
\graphicspath{ {./images/} }

\usepackage{color}%

\fussy

\newtheorem{definition}{Definition}
\newtheorem{theorem}{Theorem}
\newtheorem{assumption}{Assumption}
\newtheorem{condition}{Condition}
\newtheorem{cor}{Corollary}

\newtheorem{lemma}{Lemma}
\theoremstyle{remark}
\newtheorem{remark}{Remark}
\IEEEoverridecommandlockouts

\allowdisplaybreaks

\begin{document}
\sloppy
\title{Zero-Delay Lossy Coding of Linear Vector Markov Sources: Optimality of Stationary Codes and Near Optimality of Finite Memory Codes
	\thanks{The authors are with the Department of Mathematics and
		Statistics, Queen's University, Kingston, Ontario, Canada, K7L
		3N6.  Email: 18mg16@queensu.ca, tamas.linder@queensu.ca,
                yuksel@queensu.ca. This work was presented in part at the 2020
                IEEE Conference on Decision and Control \cite{GhLiYu20}. This research was supported in part by the Natural Sciences and Engineering Research Council (NSERC) of Canada. 
 }
}
\author{Meysam Ghomi, Tam{\'a}s Linder, and Serdar Y{\"u}ksel}
\maketitle

\begin{abstract}
Optimal zero-delay coding (quantization) of $\mathbb{R}^d$-valued linearly
generated Markov sources is studied under quadratic distortion. The structure and existence of
deterministic and stationary coding policies that are optimal for the infinite horizon average
cost (distortion) problem are established. Prior results studying the optimality
of zero-delay codes for Markov sources for infinite horizons either considered finite
alphabet sources or, for the $\mathbb{R}^d$-valued case, only showed the
existence of deterministic and non-stationary Markov coding policies or those
which are randomized. In addition to existence results, for finite blocklength
(horizon) $T$ the performance of an optimal coding policy is shown to approach
the infinite time horizon optimum at a rate $O(\frac{1}{T})$. This gives an
explicit rate of convergence that quantifies the near-optimality of finite window (finite-memory)
codes among all optimal zero-delay codes. 
\end{abstract}
	
	\begin{IEEEkeywords}
Quantization, Zero-Delay Coding, Networked Control Systems
\end{IEEEkeywords}

\section{Introduction}

In time-sensitive applications (such as networked control systems), causality in
encoding and decoding is a natural limitation. With this motivation, in this
paper we consider optimal zero-delay lossy coding for $\mathbb{R}^d$-valued Markov 
sources. In the zero-delay coding problem, the encoder encodes a source without
delay and transmits it to a decoder which  also operates without
delay.
	
We assume that the source $\{X_t \}_{t\geq0}$ is a time-homogenous
$\mathbb{R}^d$-valued discrete-time Markov process. For such a process, the
distribution of $\{X_t \}_{t\geq0}$ is uniquely determined by the initial
distribution $\pi_0$ (i.e., the distribution of $X_0$) and the transition kernel
$P(dx_{t+1}|x_t )$.

The encoder encodes (quantizes) the source samples and transmits the encoded
versions to a receiver over a discrete noiseless channel with finite input and
output alphabet $\mathcal{M} :=\{1, 2, \ldots, M\}$. The encoder is defined by a
coding policy $\Pi$, which is a sequence of Borel measurable functions
$\{\eta_t\}_{t\geq0}$ with
$\eta_t : \mathcal{M}^t \times (\mathbb{R}^d )^{t+1} \rightarrow
\mathcal{M}$. At time $t$, the encoder transmits the $\mathcal{M}$-valued
message
\begin{equation*}
  q_t=\eta_t(I_t)
\end{equation*}
where $ I_0 = X_0, I_t = (q_{[0,t-1]}, X_{[0,t]})$ for $t \geq 1$. Throughout
the paper we use the notation $q_{[0,t-1]} = (q_0,\ldots , q_{t-1})$ and
$X_{[0,t]} = (X_0, X_1, . . . , X_t )$. The set of admissible coding policies,
denoted by ${\bf \Pi}_A$, is the collection of all such zero-delay 
policies. Note that for fixed $q_{[0,t-1]}$ and $X_{[0,t-1]}$, as a function of
$X_t$, the encoder $\eta_t(q_{[0,t-1]}, X_{[0,t-1]},\, \cdot\,)$ is a Borel
measurable mapping of $\mathbb{R}^d$ into the finite set
$\mathcal{M}$. Therefore, at each time $t \geq 0$, as noted in
\cite{YukLinZeroDelay}, the coding policy selects a quantizer
$Q_t : \mathbb{R}^d \rightarrow \mathcal{M}$ based on past information
$(q_{[0,t-1]}, X_{[0,t-1]})$, and then quantizes $X_t$ as $q_t = Q_t (X_t
)$. Because of this, we refer to $\Pi$ as a quantization policy. 
	
The decoder without any delay generates the reconstruction $U_t$ using decoder
policy $\gamma = \{\gamma_t\}_{t\geq0}$, where the
$\gamma_t : \mathcal{M}^{t+1} \rightarrow \mathcal{U}$, are measurable functions
for $t\geq0$, with $\mathcal{U} \subset \mathbb{R}^d$ being the reconstruction
alphabet. Thus $U_t$ is given by
	\begin{equation*}
	U_t = \gamma_t (q_{[0,t ]}).
	\end{equation*}

In the finite horizon problem the goal is to minimize the average cumulative
cost (distortion) for a time horizon $T\in \mathbb{N}$ given by
\begin{equation}\label{finite}
  J(\pi_0,\Pi, \gamma, T ) \coloneqq \boldsymbol{E}_{\pi_0}^{\Pi,\gamma} \left[
    \frac{1}{T}\sum_{t=0}^{T-1}c_0(X_t,U_t) \right],  
\end{equation}
over the set of all admissible policies ${\bf \Pi}_A$, where
$c_0 : \mathbb{R}^d \times \mathcal{U} \rightarrow \mathbb{R}$ is a nonnegative
Borel measurable cost function (distortion measure) and
$\boldsymbol{E}_{\pi_0}^{\Pi,\gamma}$ denotes expectation with initial
distribution $\pi_0$ for $X_0$, under the quantization policy $\Pi$ and receiver
policy $\gamma$.

In the infinite horizon problem, the goal is to minimize the long-term average
cost (distortion) given by
\[
 J(\pi_0,\Pi,\gamma) \coloneqq \limsup_{T\to \infty}  \boldsymbol{E}_{\pi_0}^{\Pi,\gamma} \left[
    \frac{1}{T}\sum_{t=0}^{T-1}c_0(X_t,U_t) \right],  
\]
over all admissible policies.         
     
     \subsection{Brief literature review and contributions}
   
Two important structural results for the finite horizon problem (\ref{finite})
have been developed by Witsenhausen \cite{Witsenhausen}, and Walrand and Varaiya
\cite{WalrandVaraiya}. These results are stated in the following two
theorems. We adopt the presentation given in \cite{YukLinZeroDelay}.
	
\begin{theorem}\cite{Witsenhausen}
  For the finite horizon problem, any zero-delay quantization policy
  $\Pi = \{\eta_t\}$ can be replaced, without any loss in performance, by a
  policy $\hat{\Pi}= \{\hat{\eta}_t\}$ which only uses $q_{[0,t-1]}$ and $X_t$
  to generate $q_t$, i.e., such that $q_t = \hat{\eta}_t (q_{[0,t-1]}, X_t )$
  for all $t = 1,\ldots, T-1$.
\end{theorem}

For a complete, separable, and metric (Polish) space $\mathcal{X}$ and its Borel
sets $\mathcal{B}(\mathcal{X})$, let $\mathcal{P}(\mathcal{X})$ denote the space
of probability measures on $(\mathcal{X},\mathcal{B}(\mathcal{X}))$ equipped with the
topology of weak convergence. Given a quantization policy $\Pi$, for all
$t \geq 1$ let $\pi_t \in \mathcal{P}(\mathbb{R}^d) $ be the regular conditional
probability defined by
\begin{equation*} \label{pi_t}
	\pi_t(A)\coloneqq P(X_t \in A | q_{[0,t-1]})
\end{equation*}
for any Borel set $A \in \mathcal{B}(\mathbb{R}^d)$.
	
The following result is by Walrand and Varaiya \cite{WalrandVaraiya} where
finite-alphabet sources were studied. In \cite{YukIT2010arXiv} this
result was extended to the more general case of $\mathbb{R}^d$-valued sources.
\begin{theorem}\label{thm2}\cite{WalrandVaraiya}\cite{YukIT2010arXiv}
  For the finite horizon problem, any zero-delay quantization policy can be
  replaced, without loss in performance, by a policy which at any time
  $t = 1,\ldots, T-1$ only uses the conditional probability measure
  $\pi_t = P(dx_t |q_{[0,t-1]})$ and  $X_t$ to generate $q_t$. In other
  words, at time $t$ such a policy $\hat{\eta}_t$ uses $\pi_t$ to select a
  quantizer $Q_t = \hat{\eta}(\pi_t )$, where
  $Q_t : \mathbb{R}^d \rightarrow \mathcal{M}$, and then $q_t$ is generated as
  $q_t = Q_t (X_t )$.
\end{theorem}

We call a policy of the type in Theorem 2 a Walrand-Varaiya-type
policy. Such a policy is also called a Markov coding policy.  In the literature
several results related to zero delay coding and causal coding are
available. Notably, \cite{NeuhoffGilbert} and \cite{AsnaniWeissman} consider
causal lossy source coding where the reconstruction of the present source sample
is restricted to be a function of the present and past source samples, while the
code stream itself may be non-causal and have variable
rate. In \cite{NeuhoffGilbert} it was shown that for memoryless sources, causal source
coding cannot achieve any of the vector quantization advantages. In addition,
\cite{NeuhoffGilbert} also showed that for stationary memoryless sources, an
optimal causal coder can be replaced by one that time shares between at most two
memoryless coders, without loss in performance. In \cite{LinderZamir}, results
on causal coding by Neuhoff and Gilbert are extended to (stationary) sources
with memory, under high resolution conditions for mean squared error distortion.

Structural results for the finite horizon coding problem have been developed in
a number of  papers. As mentioned before, the classic works by
Witsenhausen \cite{Witsenhausen} and Walrand and Varaiya \cite{WalrandVaraiya},
which use two different approaches, are of particular relevance.  An extension
to the more general setting of non feedback communication was given by
Teneketzis \cite{Teneketzis}, and \cite{YukIT2010arXiv} also extended these
results to more general state spaces; see also \cite{YukLinZeroDelay} and
\cite{YukselBasarBook} for a more detailed overview. Optimal zero delay coding of
Markov sources over noisy channels without feedback was considered in
\cite{Teneketzis} and \cite{MahTen09}. We refer to
\cite{kaspi2012structure, kaspi2014zero,NayyarTeneketzis} for further results on
zero-delay or causal coding in  multi-user systems.
	
In this paper we also investigate how fast the optimum finite blocklength (time
horizon) distortion converges to the optimum (infinite horizon) distortion. An
analog of this problem in block coding is the speed of convergence of the finite
block length encoding performance to Shannon's distortion rate function. For
stationary and memoryless sources, this speed of convergence was shown to be of
the type $O\big(\frac{\log T}{T}\big)$ \cite{pilc1967coding},
\cite{zhang1997redundancy}. See also \cite{kostina2012fixed} for a detailed
literature review and further finite blocklength performance bounds. 

	
A large body of work involves convex analytic or information theoretic relaxation of the
operational problem presented above, where the constraint on the number of bits
is replaced with entropy (which may replace the fixed-rate with variable-rate
constraints) or mutual information constraints (which has a more relaxed,
Shannon theoretic infinite-dimensional, interpretation); see \cite[Section 5.4]{YukselBasarBook} for a detailed discussion. In this case, the
analysis often relies on deriving lower bounds and upper bounds on the optimal
performance, or establishing asymptotic tightness conditions. 

For lower bounds, primary methods build on Shannon lower bounding techniques (and the Gaussian measure's extremal properties), entropy-power inequality based bounds, or a {\it sequential-rate distortion theoretic} formulation where the minimization of directed mutual information is performed over causal kernels as in \cite{gorbunov1973nonanticipatory}, and which has been investigated further in a series of recent publications including \cite{tanaka2015semidefinite,stavrou2018zero,derpich2012improved,stavrou2018optimal,stavrou2021asymptotic,tanaka2016semidefinite,tatikonda2004stochastic,charalambous2014nonanticipative,kostina2019rate,stavrou2018fixed,kostina2017data}.

Related to the above, when an actual channel is present, using
channel-source coding separation based methods via the rate-distortion
function and Shannon capacity dualities also leads to useful bounds.
Perhaps the earliest papers giving such formulations are \cite{Elias},
\cite{Goblick} and \cite{Ziv}. Transmission over scalar Gaussian
channels has been also studied in \cite{Elias}, \cite{SK} and
\cite{GallagerNakiboglu}, where the error exponents were shown to be
unbounded (and the error probability was shown to decrease at least
doubly exponentially in the block-length). Transmission of linear
Gaussian sources over Gaussian channels ({\it a matched pair}, in the
sense of rate-distortion achieving and capacity achieving properties
of Gaussian models), in the scalar setup was considered in
\cite{banbas87b}, \cite{banbas89}, where the latter arrived at
tightness of information theoretic inequalities; this result has been
re-discovered later but also with some generalizations (e.g.,
  \cite{stavrou2021lqg} is a recent work considering linear systems
  and Gaussian channels in the presence of side information).

For upper bounds, methods based on high-rate quantization (and the corresponding
uniform quantization and space filling analysis), dithering (allowing for
uniformization), and entropy-power inequalities (further refining Gaussian based
bounds) have been studied; see e.g., \cite{zamir1992universal}, \cite{Silva1}, \cite{kostina2019rate}, \cite{ostergaard2021stabilizing} \cite{silva2015characterization} \cite{stavrou2018zero})

In this paper we study linear Markovian systems driven by noise and consider the
quadratic cost (mean squared distortion). Even though such systems are likely
the most important and commonly adopted ones in applications (in systems and
control theory, signal processing, and in estimation theory), their analysis in
the context of zero-delay coding are quite challenging since the costs are not
bounded. Accordingly, we will develop a number of results to address these
technical challenges. To make the presentation accessible, many of the technical
results will be presented in the appendix.

\medskip

\noindent \textit{Contributions:}

We assume that the $\mathbb{R}^d$-valued source is a linearly generated stable
Markov process and consider zero-delay quantization policies where the
quantizers have convex codecells and the cost function is the squared
distortion. Under these assumptions, our main result, Theorem \ref{thm1Lin},
demonstrates the existence of globally optimal \emph{deterministic} and
\emph{time-invariant} (stationary) Markov policies. In addition, we also show
that the (optimum) performance of such a policy for a finite time horizon $T$
converges to the infinite-horizon optimal performance at a
rate~$O(\frac{1}{T})$.

The following papers studying the infinite horizon average cost optimality in
the fixed-rate zero-delay quantization are most relevant to our work:

\begin{itemize}
\item In \cite{BorkarMitterTatikonda} a formulation for optimal average-cost
  zero-delay coding as an infinite horizon optimal stochastic control problem
  was introduced; this formulation has been an inspiration for our analysis. In
  particular, in \cite{BorkarMitterTatikonda} a stochastic control formulation
  of zero-delay quantization was given under more restrictive assumption than in
  this paper: the set of admissible quantizers in that paper was restricted to the set
  of nearest neighbor quantizers, and other conditions were placed on the
  dynamics of the system.  In contrast, we impose the more relaxed assumption
  that the quantizers have convex codecells (this class of quantizers includes
  the set of nearest neighbor quantizers).  Furthermore the proof technique used
  in \cite{BorkarMitterTatikonda} relies on the fact that the source is
  partially observed unlike in our case. As noted in \cite{YukIT2010arXiv}, for
  the partially observed case, the structure of the encoder decoder pairs
  considered in \cite{BorkarMitterTatikonda} is suboptimal since the
  measurements are not Markovian.

\item In \cite{wood2016optimal}, the source was assumed to have finite alphabet;
  however, in our case the source is taking values in $\mathbb{R}^d$ and in this
  sense the present paper generalizes \cite{wood2016optimal} to the
  technically more demanding continuous source case.  On the
  other hand, \cite{wood2016optimal}  established a global optimality
  result with no restrictions on the structure of quantizers. Here, we impose
  codecell convexity for technical reasons. 

\item Finally, in \cite{YukLinZeroDelay} only the optimality of
  deterministic and \emph{non-stationary} encoding policies, or of \emph{randomized and
  stationary} policies were established, and here we prove the optimality and existence of
  \emph{stationary and deterministic} quantization policies  and also obtain
  convergence rates for finite-memory codes, thereby generalizing
  \cite{YukLinZeroDelay} in these two aspects.
\end{itemize}

The paper is organized as follows. In Section II, after reviewing some
definitions we transform the problem into Markov decision process (MDP)
framework, and we provide some preliminary results. The main result, Theorem~\ref{thm1Lin},
is presented in Section III; to prove it we consider the discounted infinite
horizon problem followed by infinite horizon average cost problem and the proof
of Theorem~\ref{thm1Lin}. Some background material on MDPs along with useful
lemmas and theorems are presented in Appendix~A. Finally, some proofs are
relegated to Appendix~B.

\section{Preliminaries and Some Supporting Results}

In this section we present some properties of quantizers, from a different
viewpoint than is usual in source coding, that will be important in
the sequel.


A sequence of probability
measures $\{\mu_n\}$ on $\mathbb{R}^d$ is said to converge  \textit{weakly} to
$\mu \in \mathcal{P}(\mathbb{R}^d)$  if for every continuous and bounded
$f: \mathbb{R}^d \rightarrow \mathbb{R}$, we have
$\int_{\mathbb{R}^d} f(x)\mu_n(dx) \rightarrow \int_{\mathbb{R}^d} f(x)\mu(dx)$.
	
For $\mu,\nu \in \mathcal{P}(\mathbb{R}^d)$, the \textit{total variation} metric
is defined as
\begin{align}
 \|\mu - \nu\|_{TV}= \sup_{g:\|g\|_{\infty}\leq 1} \left|\int_{\mathbb{R}^d} g(x)\mu(dx) - \int_{\mathbb{R}^d}
  g(x)\nu(dx)\right|
  \label{eqtotvar} 
\end{align}
where the supremum is over all measurable real $g$ such that $\|g\|_{\infty}=
\sup\limits_{x\in \mathbb{R}^d} |g(x)| \leq 1$.

\begin{definition}\cite{villani2008optimal}
The space of probability measures with finite second moment is
\begin{equation*}
\mathcal{P}_2(\mathbb{R}^d)\coloneqq\{\mu \in \mathcal{P}(\mathbb{R}^d) : \int
\|x\|^2 \mu(dx) < \infty \}
\end{equation*}
 where $\|\cdot\|$ denotes the the Euclidean ($l_2$) norm.
\end{definition}

\begin{definition}\cite{villani2008optimal}
The order-2 Wasserstein distance for two probability distributions $\mu, \nu \in \mathcal{P}_2(\mathbb{R}^d)$ is defined as
\begin{equation*}
	\rho_2(\mu,\nu) = \inf_{\lambda \in \mathcal{H}(\mu \times \nu)}
        \left(\int_{\mathbb{R}^d \times \mathbb{R}^d} \|x-y\|^2
          \lambda(dx,dy)\right)^{\frac{1}{2}}, 
\end{equation*}
where $\mathcal{H}(\mu \times \nu)$ denotes the set of probability measures on
$\mathbb{R}^d \times \mathbb{R}^d$ with first marginal $\mu$ and second marginal
$\nu$.
\end{definition}
	
For compact subsets of $\mathbb{R}^d$, the Wasserstein distance of order $2$
metrizes the weak topology on the set of probability measures on $\mathbb{R}^d$
(see \cite[Theorem 6.9]{villani2008optimal}). For non-compact subsets, weak
convergence combined with convergence of second moments (that is of
$\int \|x\|^2 \mu_n(dx) \to \int \|x\|^2 \mu(dx)$) is equivalent to convergence
in order-2 Wasserstein distance.
	
\begin{definition}
  An $M$-cell quantizer $Q$ on $\mathbb{R}^d$ is a (Borel) measurable mapping
  $Q : \mathbb{R}^d \rightarrow \mathcal{M}$. We let $\mathcal{Q}$ denote the
  collection of all $M$-cell quantizers on $\mathbb{R}^d$.
\end{definition}

Observe that each $Q \in \mathcal{Q}$ is uniquely characterized by its
quantization cells (or bins)
$B_i = Q^{-1}(i ) = \{x : Q(x) = i \}, i = 1,\ldots, M$ which form a
measurable partition of $\mathbb{R}^d$.

\begin{definition}\label{defPiA} An (admissible) quantization policy
  $\Pi = \{\eta_t \}_{t \geq 0}$ belongs to ${\bf \Pi}_W$ (i.e., it is a
  Walrand-Varaiya type policy) if there exist a sequence of
  mappings $\{ \hat{\eta}_t \}$ of the type
  $\hat{\eta}_t : \mathcal{P}(\mathbb{R}^d ) \rightarrow \mathcal{Q}$ such that
  for $Q_t = \hat{\eta}_t(\pi_t )$ we have $q_t = Q_t (X_t ) = \eta_t (I_t )$.
\end{definition}

Suppose we use a quantizer policy $\Pi=\{\hat{\eta}_t\} \in {\bf \Pi}_W$. Then, using
standard properties of conditional probability, building on
\cite{YukLinZeroDelay} we can obtain the following filtering equation for the
evolution of $\pi_t$:
\begin{align}
  \pi_{t+1}(dx_{t+1})&=\frac{P(dx_{t+1},q_t | q_{[0,t-1]})}{P(q_t | q_{[0,t-1]})} \nonumber \\ &= \frac{\int_{\mathbb{R}^d}\pi_t(dx_t)P(q_t|\pi_t,x_t)P(dx_{t+1}|x_t)}{\int_{\mathbb{R}^d}\int_{\mathbb{R}^d}\pi_t(dx_t)P(q_t|\pi_t,x_t)P(dx_{t+1}|x_t)} \nonumber \\
                     &=\frac{1}{\pi_t(Q_t^{-1}(q_t))}\int_{Q_t^{-1}(q_t)}\pi_t(dx_t)P(dx_{t+1}|x_t).
                       \label{eqfil}   
\end{align}
Thus $\pi_{t+1}$ depends only on $\pi_t , Q_t,$ and $q_t$, which implies that
$\pi_{t+1}$ is conditionally independent of $(\pi_{[0,t-1]}, Q_{[0,t-1]})$ given
$\pi_t$ and $Q_t$ . Thus, $\{\pi_t \}$ can be viewed as
$\mathcal{P}(\mathbb{R}^d )$-valued controlled Markov process \cite{HeLa96} (see
also Appendix~A), with $\mathcal{Q}$-valued
control $\{Q_t \}$ having transition kernel $P(d\pi'|\pi,Q)$ determined by 
\eqref{eqfil}. The average cost up to time $T - 1$ is given by (see also
\cite{YukLinZeroDelay}) 
\begin{equation}\label{finiteAC}
\boldsymbol{E}_{\pi_0}^{\Pi} \left[ \frac{1}{T}\sum_{t=0}^{T-1}c(\pi_t,Q_t)
\right]=\inf_{\gamma} J(\pi_0,\Pi,\gamma,T), 
\end{equation}
where
\begin{equation}\label{eqC}
c(\pi_t,Q_t)\coloneqq\sum_{i=1}^{M}\inf_{u\in \mathcal{U}}\int_{Q^{-1}_t(i)} \pi_t(dx)c_0(x,u).
\end{equation}

For the mean squared distortion  $c_0(x,u)=\|x-u\|^2$  (which is our focus),  the optimum receiver
$\gamma_t$ at
time $t$ is
explicitly given by
\begin{equation}
 \label{eq:optdec}
\gamma_t(i) = \frac{1}{\pi(Q^{-1}_t(i))}\int_{Q^{-1}_t(i)} x \pi_t(dx), \quad
i=1,\ldots,M. 
\end{equation}
	
\begin{definition} \cite{YukLinZeroDelay} \label{sdef} Let $\mathcal{G}$ denote
  the set of all probability measures on $\mathbb{R}^d$ admitting densities that
  are bounded by $C$ and Lipschitz with constant $C_1$.
\end{definition}
	
In \cite[Lemma 3]{YukLinZeroDelay} it is shown that $\mathcal{G}$ is closed in
$\mathcal{P}(\mathbb{R}^d)$. Note that $\mathcal{G}$ is also closed in
$\mathcal{P}_2(\mathbb{R}^d)$, since the Wasserstein convergence is stronger
than the weak convergence. Let
${\cal Z}\coloneqq {\cal G} \cap {\cal P}_2(\mathbb{R}^d)$, be the intersection
of ${\cal G}$ and ${\cal P}_2(\mathbb{R}^d)$. 

\begin{remark} \label{rem1}
  Due to our assumptions on the source $\{X_t\}$ (see Section~III), the
  distribution of $X_{t+1}$ has (conditional) density function $\phi(\cdot|x)$
  given $X_t=x$,
  which is positive everywhere, bounded, and Lipschitz uniformly in~$x$.  Thus
  (with appropriate constants $C$ and $C_1$), $P(dx_{t+1}|x_t) \in \mathcal{G}$
  for all $x_t\in \mathbb{R}^d$ and thus the filtering equation \eqref{eqfil}
  implies that under any policy $\Pi\in {\bf \Pi}_W$, we have  $\pi_t\in \mathcal{G}$ for all $t\ge 0$ if
  $\pi_0\in \mathcal{G}$. The assumptions on the source will also imply that
  $\pi_t$ has finite second moment (with probability one) for all $t\ge 0$ if $\pi_0$ has finite second
  moment (see \eqref{eq3Moment}), so we obtain $\pi_t\in \mathcal{Z}$ for all
  $t\ge 0$ if $\pi_0\in \mathcal{Z}$. Thus we can make $\mathcal{Z}$  the state space of
  our Markov decision process. 
\end{remark}

From now on, we restrict the set of quantizers to quantizers having  convex cells
\cite{YukLinZeroDelay}. Formally, this quantizer class $\mathcal{Q}_c$ is
defined by
\begin{equation*}
\mathcal{Q}_c=\{Q \in \mathcal{Q} : Q^{-1}(i) \subset \mathbb{R}^d \; \text{is convex for} \\ \; i=1,\ldots,M\}. 
\end{equation*} 
Thus, we replace $\mathcal{Q}$ with $\mathcal{Q}_c$ in Definition
\ref{defPiA} to obtain the new class of policies denoted by ${\bf
  \Pi}^C_W$. 

\begin{definition}  We denote by   ${\bf \Pi}^C_W$ the set of all quantization policies $\Pi= \{\hat{\eta}_t\}
  \in {\bf \Pi}_W$ such that $\hat{\eta}_t : \mathcal{P}(\mathbb{R}^d ) \rightarrow
  \mathcal{Q}_c$, i.e., $Q_t = \hat{\eta}_t(\pi_t ) \in \mathcal{Q}_c$ for all
  $t \geq 0$. Furthermore,  ${\bf \Pi}^C_{WS}$ denotes the set of all quantization policies in ${\bf \Pi}^C_W$
  that are stationary, i.e., the policy $\{\hat{\eta}_t\}$ does not depend on
  the time index $t$.
\end{definition}

\begin{remark}   \mbox{\ } \\ \mbox{\ } \vspace{-15pt}
\begin{itemize}     
  \item[(i)] The set ${\bf \Pi}^C_W$ is called the set of Markov quantization policies
    and ${\bf \Pi}^C_{WS}$ is called the set of stationary Markov quantization
    policies.
  \item[(ii)] The convex codecell restriction may lead to
    suboptimality in general; however it includes the class of nearest
    neighbor quantizers studied in \cite{BorkarMitterTatikonda}. For
    multiresolution scalar quantizers (MRSQ) and the squared error
    distortion measure,
    \cite{antos2012codecell,muresan2008quantization} showed that for
    discrete and continuous sources (even with bounded continuous
    densities), optimal fixed rate multiresolution scalar quantizers
    cannot have only convex codecells, proving that the convex
    codecells assumption leads to a loss in performance. We introduce the convex codecell assumption for
      technical reasons; without this assumption the analysis of
      recursive policies seems very hard. Indeed,  the parametric
      representation of convex codecell quantizers allowed
      \cite{YukselOptimizationofChannels} to establish compactness and
      desired convergence properties. In particular, in the absence of
      such a condition, it was shown in
      \cite[p.~878]{YukselOptimizationofChannels} that the space of
      quantizers is not closed under weak convergence.
  \end{itemize} 
\end{remark}
	
Following \cite{YukLinZeroDelay} and \cite{YukselOptimizationofChannels}, in
order to facilitate the stochastic control analysis of the quantization problem
we will use an alternative representation of quantizers. A quantizer
$Q : \mathbb{R}^d \rightarrow \mathcal{M}$ with cells $\{B_1, \ldots, B_M\}$,
can also be identified with the stochastic kernel (regular conditional
probability) on $\mathcal{M}$ given $\mathbb{R}^d$, also denoted by $Q$, defined by
\begin{equation*}
	Q(i|x)=1_{\{x \in B_i \}}, \quad i =1,\ldots,M.
\end{equation*}

As in \cite{YukselOptimizationofChannels,YukLinZeroDelay}, we say that a
sequence of quantizers $Q_n$ converges to $Q$ at $P \in {\cal P}(\mathbb{R}^d)$
if $PQ_n \to PQ$, where $\mu \bar{Q}$ denotes  the probability measure on
$\mathbb{R}^d\times \mathcal{M}$ induced by a  $\mu \in \mathcal{P}(\mathbb{R}^d)$
and a conditional probability $\bar{Q}$ on $\mathcal{M}$ given
$\mathbb{R}^d$. Here we consider convergence in the order-2 Wasserstein
distance. We note that by \cite[Lemma~2]{YukLinZeroDelay}, the convergence of quantizers with convex codecells holds 
simultaneously for all admissible input probability measures in ${\cal Z}$ and
accordingly we will not need to specify $P$ explicitly.

The following lemma shows the compactness of $\mathcal{Q}_c$ in the order-2 Wasserstein
topology. The proof is given in Appendix B.
\begin{lemma} \label{qccompact} 
	$\mathcal{Q}_c$ is compact in the order-2 Wasserstein topology at any input $P \in {\cal Z}$. 
\end{lemma}

From now on we assume the mean squared distortion $c_0(x,u)=\|x-u\|^2$.  The
following lemmas are proved in Appendix B.

\begin{lemma}\label{ctsc}
  The cost function $c(\pi,Q)$ is lower semi-continuous in $(\pi,Q)$, that is,
  when $(\pi_n,Q_n) \rightarrow (\pi,Q)$ (in order-2 Wasserstein distance), 
  then $$\liminf_{n \rightarrow \infty} c(\pi_n,Q_n) \geq c(\pi,Q).$$
Also, $c(\pi,Q)$ is continuous in $Q$ for every fixed $\pi \in {\cal Z}$, i.e.,
if $Q_n \rightarrow Q$, then  $ c(\pi,Q_n) \to   c(\pi,Q)$.
\end{lemma}

Recall the transition probability $P(d\pi'|\pi,Q)$ of our MDP determined by the
filtering equation \eqref{eqfil}.

\begin{lemma}\label{ctsKernel}
The function $Pg(\pi,Q)\coloneqq\int_{\mathcal{P}_2(\mathbb{R}^d)\times
  \mathcal{Q}_c} g(\pi')P(d\pi'|\pi,Q)$ is continuous in $(\pi,Q)$ (i.e. is
continuous when $(\pi_n,Q_n) \rightarrow (\pi,Q)$ in order-2 Wasserstein
distance on $\mathcal{Z}\times \mathcal{Q}_c$), for every continuous bounded
function $g: {\cal Z} \to \mathbb{R}$.  Moreover, for any fixed $\pi\in
\mathcal{Z}$, $Pg(\pi,Q) $ is continuous in $Q\in \mathcal{Q}_c$ for any
continuous function $g$. 
\end{lemma}
	
In the theory of Markov decision processes (MDPs) (see also Appendix~A),  the
so-called \textit{measurable selection condition} (\cite[Assumption
3.3.1]{HeLa96}) guarantees the measurability of the value function and existence
of a minimizer (\cite[Theorem~3.3.5]{HeLa96}). The following assumption, which is
stated for the Markov control model of our zero-delay quantization setup, is
referred to as the \textit{measurable selection condition}.

\begin{assumption}\cite[Assumption 3.3.1]{HeLa96}\label{measurableSelectionCondition}
  The Markov control model and a given measurable function $u: {\cal Z} \to
  \mathbb{R}$ are such that $u^*: \mathcal{Z} \to \mathbb{R}$ defined by 
  \begin{equation*}
		u^*(\pi) \coloneqq \inf_{Q\in \mathcal{Q}_c} \left(c(\pi,Q) + \int_{{\cal Z}} u(\pi')P(d\pi'|\pi,Q)\right), \quad \pi \in {\cal Z}
\end{equation*}
	is measurable and there exist a measurable $\hat{\eta} : \mathcal{Z}\to
        \mathcal{Q}_c$ such that for any $\pi\in \mathcal{Z}$,  $Q= \hat{\eta}(\pi)$ attains the minimum 
at $\pi$, i.e., 
\begin{equation}\label{u8}
		u^*(\pi)=c(\pi,\hat{\eta}(\pi)) + \int_{{\cal Z}} u(\pi')P(d\pi'|\pi,\hat{\eta}(\pi)),
                \quad \text{for all } \pi \in {\cal Z}. 
\end{equation} 
\end{assumption}

The following is a sufficient condition for the Assumption~\ref{measurableSelectionCondition} to hold. Note that conditions (i)-(iii) hold
in our setting by Lemmas \ref{qccompact}, \ref{ctsc}, and
\ref{ctsKernel}. Therefore Theorem~\ref{measurableselectionThm} below holds for
our model.

\begin{condition}\label{condition1}
\mbox{}
\begin{itemize}
	\item[(i)] The quantizer space (i.e. action space) $\mathcal{Q}_c$ is compact for every fixed $\pi$.
	\item[(ii)] The one-stage cost function $c(\pi,Q)$ is lower semi-continuous in $(\pi,Q)$.
	\item[(iii)] The transition kernel $P$ is such that 
\begin{equation*}
Pg(\pi,Q)\coloneqq\int_{\mathcal{P}_2(\mathbb{R}^d)\times \mathcal{Q}_c} g(\pi')P(d\pi'|\pi,Q)
\end{equation*}
is continuous in $(\pi,Q)$ for every continuous and bounded $g$ on ${\cal Z}$.
\end{itemize}
\end{condition}

\begin{theorem}\cite[Theorem 3.3.5]{HeLa96} \label{measurableselectionThm}
Condition \ref{condition1} implies Assumption \ref{measurableSelectionCondition}
for any nonnegative measurable $u:\mathcal{Z}\to \mathbb{R}$. Moreover if, $u$ is nonnegative and lower semi-continuous then the function $u^*$ in (\ref{u8}) is lower semi-continuous.
\end{theorem}

\section{Infinite Horizon Problem of Linear Systems under Quadratic Cost}
We consider the linear system given in the following assumption.
	
\begin{assumption}\label{assumption3}
The source  $\{X_t\}$ can be expressed in the linear stochastic realization form
\begin{equation}\label{LQG}
	X_{t+1}=AX_t+W_t,
\end{equation}
where $A$ is a $d\times d$ real matrix and ${W_t}$ is an independent and
identically distributed (i.i.d.) vector noise sequence which is independent of
$X_0$. Moreover, assume the following:
\begin{itemize}
\item[(i)]	The maximum singular value of $A$, denoted by $\alpha$, is less than 1 (i.e. maximum eigenvalue of the matrix $A'A$ is less than 1, where $A'$ is the transpose of the matrix $A$).
\item[(ii)] $\mathcal{U}=\mathbb{R}^d$.
\item[(iii)] The cost for the pair $(x,u)$ is given by $c_0(x,u)=\|x-u\|^2$.
\item[(iv)] The $W_t$ have a common probability density function $\varphi$ that is
  positive, bounded,  and Lipschitz continuous. 
\item[(v)] $\sigma^2\coloneqq\boldsymbol{E}[\|W_t\|^2] < \infty$.
\item[(vi)] The initial distribution $\pi_0$ for $X_0$ admits a density such that $\boldsymbol{E}_{\pi_0}[\|X\|^2]<\infty$ or it is a point mass $\pi_0=\delta_{x_0}$.
\end{itemize}
\end{assumption}
Note that assumption~(iv) implies that for each fixed $x \in \mathbb{R}^d$, the distribution
of $X_{t+1}$, (i.e. $Ax+W_t$), has (conditional) density function
$\phi(\cdot|x)$ which is positive everywhere,   bounded,  and Lipschitz
uniformly in~$x$. Thus (with appropriate constants $C$ and $C_1$) we have
$\phi(dy|x) \in \mathcal{G}$ for all $x\in \mathbb{R}^d$, where $\mathcal{G}\subset
\mathcal{P}(\mathbb{R}^d)$ was defined in Definition~\ref{sdef}. As we observed
in Remark~\ref{rem1}, this implies that $\pi_t\in \mathcal{Z}$ for all $t\ge
0$.

For any initial distribution $\pi_0\in \mathcal{Z}$, the long-term
(infinite-horizon) minimum cost (distortion) of a quantization policy $\Pi\in
{\bf \Pi}^C_W$ is
\begin{equation*}
J(\pi_0,\Pi) \coloneqq \limsup_{T\to \infty} \boldsymbol{E}_{\pi_0}^{\Pi}
\left[\frac{1}{T}  \sum_{t=0}^{T-1} c(\pi_t,Q_t) \right]
\end{equation*}
and the optimal cost over all policies in ${\bf \Pi}^C_W$  is
\[
J(\pi_0)  \coloneqq  \inf_{\Pi \in {\bf \Pi}^C_W} J(\pi_0,\Pi). 
\]

Our main result is the following theorem.

\begin{theorem}\label{thm1Lin}   \mbox{\ } \\ \mbox{\ } \vspace{-15pt}
\begin{itemize}
\item[(i)]
Under Assumption \ref{assumption3}, for any initial distribution $\pi_0$ 
\begin{equation} \label{mainthm}
J(\pi_0,\Pi^*) = \inf_{\Pi \in {\bf \Pi}^C_W} J(\pi_0,\Pi) = \min_{\Pi \in {\bf \Pi}^C_{WS}} J(\pi_0,\Pi).
\end{equation}
That is, there exists a deterministic and stationary policy
$\Pi^* \in {\bf \Pi}^C_{WS}$ that achieves the minimum above.
\item[(ii)] Furthermore, the finite horizon distortion of the optimal policy
  $\Pi$ converges to its infinite horizon distortion at a rate $O(\frac{1}{T})$;
  in particular, for all $\pi_0$ and $T \geq 1$, 
\begin{eqnarray*}
\left|\frac{1}{T}\boldsymbol{E}_{\pi_0}^{\Pi^*} \left[ \sum_{t=0}^{T-1}c(Q_t,\pi_t) \right] - J(\pi_0,\Pi^*)\right| \leq \frac{K(\pi_0)}{T},
\end{eqnarray*} 
 where $K(\pi_0) < \infty$ only depends on $\pi_0$.
\end{itemize}
\end{theorem}

We will prove the theorem in Section~\ref{subsecAv} after obtaining auxiliary
existence and optimality results for the easier-to-handle discounted cost problem in the
next section. 

	\subsection{The Discounted Cost Problem}
The discounted cost for some $\beta \in (0,1)$ and time horizon $T \ge 1 $ is defined as
\begin{equation}\label{Discfinite}
J^{\beta}(\pi_0,\Pi,T) \coloneqq \boldsymbol{E}_{\pi_0}^{\Pi} \left[ \sum_{t=0}^{T-1}\beta^tc(\pi_t,Q_t) \right], 
\end{equation}
and for the infinite horizon case, 
\begin{equation*}
J^{\beta}(\pi_0,\Pi) \coloneqq \boldsymbol{E}_{\pi_0}^{\Pi} \left[ \sum_{t=0}^{\infty}\beta^tc(\pi_t,Q_t) \right],
\end{equation*}
where $c(\pi_t,Q_t)$ is defined in (\ref{eqC}).

The goal is to find optimal policies that achieve 
\begin{equation}\label{DisOp}
J^{\beta}(\pi_0) \coloneqq  \inf_{\Pi \in {\bf \Pi}^C_{W}}J^{\beta}(\pi_0,\Pi). 
\end{equation}
We call $J^{\beta}$ the discounted \emph{value function} of the MDP. 
Let us define 
\begin{equation*}
J^{\beta}(\pi_0,T) \coloneqq \inf_{\Pi \in {\bf \Pi}^C_{W}}J^{\beta}(\pi_0,\Pi,T),
\end{equation*}
so that we have 
\begin{equation}
J^{\beta}(\pi_0) \geq \limsup_{T \rightarrow \infty} J^{\beta}(\pi_0,T).
\end{equation}
Since $J^{\beta}(\pi_0,T)$ is monotonically increasing in $T$,  the limit superior becomes a limit and thus
\begin{equation}\label{Jbetaupperbound}
J^{\beta}(\pi_0) \geq \lim_{T \rightarrow \infty} J^{\beta}(\pi_0,T).
\end{equation}


Let $v:\mathcal{Z}\to \mathbb{R}$ be lower semicontinuous and define the operator $\mathbb{H}$
by
\begin{equation}
  \label{eq:Hdef}
(\mathbb{H}v)(\pi)\coloneqq \min_{Q \in {\cal Q}_c} \left(c(\pi,Q)+\beta \int_{{\cal Z}} v(\pi_1) P(d\pi_1|\pi,Q)\right).
\end{equation}
Note that $\mathbb{H}$ indeed maps lower semicontinuous functions into lower
semicontinuous functions by
Theorem~\ref{measurableselectionThm}. 
The discounted cost optimality equation (DCOE) is defined by 
\begin{equation}\label{DCOE}
	v(\pi)=(\mathbb{H}v)(\pi), \quad \pi \in {\cal Z}.
\end{equation}

The following theorem is a version of a widely used result in the theory of Markov decision
processes.

\begin{theorem}\label{discountedThm}
	Suppose Assumption~\ref{assumption3} holds. Then, the value function $J^{\beta}(\pi_0)$ is a fixed point of the operator $\mathbb{H}$, i.e.
	\begin{equation}\label{fixedpoint}
	J^{\beta}=\mathbb{H}J^{\beta}.
	\end{equation}
	Furthermore, there exists  a deterministic stationary policy
        $\Pi=\{\hat{\eta}\} \in {\bf \Pi}^C_{WS}$ that is optimal, i.e.,
        $J^{\beta}(\pi_0) = J^{\beta}(\pi_0, \Pi)$ for all $\pi_0\in
        \mathcal{Z}$ and this policy satisfies for all $ \pi_0 \in
        {\cal Z}$,
\begin{equation}
 J^{\beta}(\pi_0) =  c(\pi_0,\hat{\eta}(\pi_0) )+\beta \int_{{\cal Z}}
 J^{\beta}(\pi') P(d\pi'|\pi_0, \hat{\eta}(\pi_0)).
 \end{equation}
\end{theorem}

Since our setup is quite non-standard, we will have to give a separate proof of
Theorem~\ref{discountedThm} after stating and proving some preliminary result. In what follows $\boldsymbol{E}_{\pi_0} [\|X_t\|^2] $
denotes the second moment of $X_t$ when $X_0\sim \pi_0$ and
$\boldsymbol{E}_{\pi_0} [\|X\|^2] = \boldsymbol{E}_{\pi_0} [\|X_0\|^2]  =
\int_{\mathbb{R}^d} \|x\|^2 \pi_0(dx)$. 

\begin{lemma}\label{JbetaLemma}
	For every initial distribution $\pi_0 \in {\cal Z}$, the value function $J^{\beta}(\pi_0)$, and hence also $J^{\beta}(\pi_0,T)$, is uniformly bounded as
	\begin{align*}
	J^{\beta}(\pi_0,T) &\leq J^{\beta}(\pi_0) \\* &\leq  \frac{1}{1-\beta} \left(\boldsymbol{E}_{\pi_0}[\|X\|^2] + \frac{1}{1-\alpha} \sigma^2\right). 
	\end{align*} 
\end{lemma}
\begin{proof}
Note that the receiver defined in \eqref{eq:optdec} minimizes $\int \|x-\gamma(Q(x))\|^2\pi(dx)$ over all $\gamma$ and hence,
\begin{equation*}
c(\pi,Q) = \min_{\gamma} \int_{\mathbb{R}^d}  \|x-\gamma(Q(x))\|^2\pi(dx) \leq \int_{\mathbb{R}^d}  \|x\|^2 \pi(dx).
\end{equation*}
On the other hand, by the properties of the process we have,
\begin{align}
&\boldsymbol{E}_{\pi_0} [\|X_{t}\|^2] \nonumber  \\ 
&= \boldsymbol{E}_{\pi_0}[\|AX_{t-1}+W_{t-1}\|^2] \nonumber  \\ 
  &=\boldsymbol{E}_{\pi_0}[X_{t-1}A'AX_{t-1}+W'_{t-1}W_{t-1}] \nonumber  \\ 
&\leq \boldsymbol{E}_{\pi_0}[\alpha\|X_{t-1}\|^2] + \sigma^2 \nonumber  \\
&\leq \boldsymbol{E}_{\pi_0}[\alpha^{t}\|X_{0}\|^2] + (\alpha^{t-1}+\ldots+\alpha+1)
                                                                            \sigma^2  \nonumber  \\
&\leq \boldsymbol{E}_{\pi_0}[\|X\|^2] + \frac{1}{1-\alpha} \sigma^2 < \infty.  \label{eq3Moment}
\end{align}
Therefore, we can bound $J^{\beta}(\pi_0)$ as
\begin{align}
J^{\beta}(\pi_0) &\leq J^{\beta}(\pi_0,\Pi) \leq
                                             \boldsymbol{E}_{\pi_0}^{\Pi}\left[\sum_{t=0}^{\infty}\beta^tc(\pi_t,Q_t)\right]
                                             \nonumber  \\*
                                           &\leq
                                                             \boldsymbol{E}_{\pi_0}^{\Pi}\left[\sum_{t=0}^{\infty}\beta^t\|X_t\|^2\right]
   \nonumber  \\ 
 &\leq \frac{1}{1-\beta} \left(\boldsymbol{E}_{\pi_0}[\|X\|^2] +
   \frac{1}{1-\alpha} \sigma^2\right).  \label{Jbetabound}
\end{align}
This together with (\ref{Jbetaupperbound}) yields the lemma.
\end{proof}

The following is a key {\it equicontinuity} lemma which is related to, but
different from, Lemma 1 in \cite{wood2016optimal}. The proof is also related to
the approach of Borkar~\cite{borkar2000average} (see also
\cite{borkar2007dynamic} and \cite{BorkarMitterTatikonda}), but our argument is
different (and more direct) since the absolute continuity conditions in
\cite{borkar2000average} are not applicable here due to quantization. As in
\cite{BorkarMitterTatikonda}, in the proof we will enlarge the space of
admissible coding policies to allow for randomization at the encoder.  Since for
a discounted infinite horizon optimal encoding problem optimal policies are
deterministic even among possibly randomized policies, allowing randomness does
not change the optimal performance.
\begin{lemma}\label{lem1Lin}
Suppose the source is generated as in (\ref{LQG}) and Assumption \ref{assumption3} holds.
Then for any initial two distributions $\mu_0,\nu_0 \in {\cal Z}$, and any $\beta \in (0, 1)$, we have
\begin{equation}
	\left|J^{\beta}(\nu_0)-J^{\beta}(\mu_0)\right| \leq
        \left(\frac{\rho_2(\nu_0,\mu_0)}{1-\alpha}+\frac{2K_1}{1-\sqrt{\alpha}}\right)\rho_2(\nu_0,\mu_0),  \label{eq:eqcontbound}
\end{equation}
where $K_1$ is a finite constant and $\rho_2(\nu_0,\mu_0)$ is the order-2
Wasserstein distance of the two initial distributions. 
\end{lemma}
	
\begin{proof}Consider the $\mathbb{R}^d \times \mathbb{R}^d$-valued process
  $\{(X_t , Y_t )\}_{t\geq 0}$ such that
  $\{X_t \}_{t\geq 0} \sim (\nu_0, P),\: \{Y_t \}_{t\geq 0} \sim (\mu_0, P),\:
  (X_0, Y_0) \sim \lambda$ where
  $\lambda \in \mathcal{P}(\mathbb{R}^d \times \mathbb{R}^d)$ with marginals
  $\nu_0$ and $\mu_0$ respectively. We further assume identical noise
  realization $W_t$ for these processes.  Assume without loss of generality that
  $J^{\beta}(\nu_0)-J^{\beta}(\mu_0) \geq 0$. Then
\begin{align*}
&	\left|J^{\beta}(\nu_0)-J^{\beta}(\mu_0)\right| = J^{\beta}(\nu_0)-J^{\beta}(\mu_0) \nonumber \\
&= \boldsymbol{E}_{\nu_0}^{\Pi_x} \left[ \sum_{t=0}^{\infty}\beta^tc_0(X_t,U_t) \right]\nonumber - \boldsymbol{E}_{\mu_0}^{\Pi_y} \left[ \sum_{t=0}^{\infty}\beta^tc_0(Y_t,\tilde{U_t}) \right],
\end{align*}
where we assume that $\Pi_x \in {\bf \Pi}^C_{W}$ and
$\Pi_y \in {\bf \Pi}^C_{W}$ achieve $J^{\beta}(\nu_0)$ and $J^{\beta}(\mu_0)$
respectively. (Note that we make this assumption only for convenience; at
this point we do not know if such optimal policies exist. However, for any
$\delta>0$ there exist $\Pi_x , \Pi_y \in {\bf \Pi}^C_{W}$  such that
$J^{\beta}(\nu_0,\Pi_x)<J^{\beta}(\nu_0)+\delta$ and
$J^{\beta}(\mu_0,\Pi_y)<J^{\beta}(\mu_0)+\delta$ and using such $\delta$-optimal
policies in the proof will
lead to the same bound as in \eqref{eq:eqcontbound}  since $\delta>0$ can be arbitrarily small.) 
		
Consider the following suboptimal encoding and decoding policy for $\{X_t\}$: The
encoder, in addition to observing the source $\{X_t\}$, has access to the noise
process $\{W_t\}$ which is independent of $X_0$. Then the encoder can generate
the source $Y_0$ through a simulation (which will be optimized later on with an
optimal Wasserstein coupling), and then produce $Y_t$ for $t \geq 0$ according to the
following equation
\begin{align}\label{eqlin}
		Y_t=A^t(Y_0-X_0)+X_t.
\end{align}
Then the encoder for $\{X_t\}$ can use the quantizer policy $\Pi_y$ and produce
the same channel symbols $\tilde{q_t}$ as $\Pi_y$ and thus the same reproduction sequence
$\tilde{U_t}=\tilde{\gamma_t}(\tilde{q}_{[0,t]})$ as the encoder and decoder for
$\{Y_t\}$. Denote this suboptimal policy by $\hat{\Pi}$. Then we get the upper
bound 
\begin{align*}
&\left|J^{\beta}(\nu_0)-J^{\beta}(\mu_0)\right| \\
&\leq \boldsymbol{E}_{\nu_0}^{\hat{\Pi}} \left[ \sum_{t=0}^{\infty}\beta^tc_0(X_t,\tilde{U_t}) \right] \nonumber - \boldsymbol{E}_{\mu_0}^{\Pi_y} \left[ \sum_{t=0}^{\infty}\beta^tc_0(Y_t,\tilde{U_t}) \right].    
\end{align*}
Since $\beta \in (0,1)$ and $c_0(x,u)=\|x-u\|^2$, we have
\begin{align}
  &\Bigl|J^{\beta}(\nu_0)-J^{\beta}(\mu_0)\Bigr|  \nonumber \\
  &\leq \sum_{t=0}^{\infty}  \boldsymbol{E}_{\lambda}\left[ \left|  X_t'X_t+2\tilde{U_t}'(Y_t-X_t)-Y_t'Y_t  \right| \right] \nonumber \\
  &= \sum_{t=0}^{\infty}  \boldsymbol{E}_{\lambda}\Bigl[ \Bigl|  X_t'X_t+2(\tilde{U_t}-Y_t+Y_t)'(Y_t-X_t) -Y_t'Y_t \Bigr| \Bigr]\nonumber \\
  &= \sum_{t=0}^{\infty}  \boldsymbol{E}_{\lambda}\left[ \left|  \|X_t-Y_t\|^2+2(\tilde{U_t}-Y_t)'(Y_t-X_t) \right| \right] \nonumber \\
  &\leq \sum_{t=0}^{\infty} \Bigl( \boldsymbol{E}_{\lambda}\Bigl[\|X_t-Y_t\|^2\Bigr] \nonumber \\
  & \quad \quad + 2\left(\boldsymbol{E}_{\lambda}\left[  \|X_t-Y_t\|^2 \right]\boldsymbol{E}_{\lambda}\left[  \|\tilde{U_t}-Y_t\|^2 \right] \right)^{\frac{1}{2}} \Bigr), \label{ineq1}
\end{align}
where the last inequality follows from the Cauchy-Schwarz inequality. 
		
Since $\tilde{U_t}$ is produced by the optimal decoder for the source $Y_t$, if we use suboptimal reconstruction $\hat{U_t}=0$ for all $t \geq 0$, we get an upper bound
		
\begin{equation*}
\boldsymbol{E}_{\lambda}\left[  \|\tilde{U_t}-Y_t\|^2 \right] \leq \boldsymbol{E}_{\lambda}\left[  \|Y_t\|^2 \right],
\end{equation*}
and moreover, (\ref{eq3Moment}) implies
\begin{align}
\boldsymbol{E}_{\lambda}\left[  \|Y_t\|^2 \right] \leq \boldsymbol{E}_{\lambda}\left[  \|Y_0\|^2 \right] + \frac{1}{1-\alpha}\sigma^2 .
\end{align}
By (\ref{eqlin}) we can write
\begin{align*}
	\|X_t-Y_t\|^2&=(X_0-Y_0)'(A^t)'A^t(X_0-Y_0),\nonumber
\end{align*}
so by the Assumption \ref{assumption3}(i), we get that 
\begin{equation*} 
	\|X_t-Y_t\|^2 \leq \alpha^t\|X_0-Y_0\|^2.
\end{equation*}
Since $\alpha < 1$, (\ref{ineq1}) gives, with $K_2\coloneqq\boldsymbol{E}_{\lambda}\left[
  \|Y_0\|^2 \right] + \frac{1}{1-\alpha}\sigma^2$,
\begin{align*}
& \Bigl|J^{\beta}(\nu_0)-J^{\beta}(\mu_0)\Bigr| \\
&\leq \boldsymbol{E}_{\lambda}\Biggl[ \sum_{t=0}^{\infty} \alpha^t\|X_0-Y_0\|^2\Biggr] \\
&+2\sum_{t=0}^{\infty}(\sqrt{\alpha})^t\sqrt{\boldsymbol{E}_{\lambda}\left[\|X_0-Y_0\|^2\right]}\sqrt{\boldsymbol{E}_{\lambda}\left[\|Y_t\|^2\right]}  \\
		&\leq \Biggl(\sum_{t=0}^{\infty} \alpha^t \Biggr)\boldsymbol{E}_{\lambda}\left[ \|X_0-Y_0\|^2\right] \\
		&\quad \quad +2\sqrt{\boldsymbol{E}_{\lambda}\left[\|X_0-Y_0\|^2\right]}\sum_{t=0}^{\infty}(\sqrt{\alpha})^t\sqrt{K_2}  \\
&=\Biggl(\frac{1}{1-\alpha}\Biggr)\boldsymbol{E}_{\lambda}\left[ \|X_0-Y_0\|^2\right] \\
		&\quad \quad +\left(\frac{2\sqrt{K_2}}{1-\sqrt{\alpha}}\right)\sqrt{\boldsymbol{E}_{\lambda}\left[\|X_0-Y_0\|^2\right]}.
\end{align*}
By the definition of the Wasserstein distance, for any $\epsilon > 0$, by
suitably choosing the joint law  $\lambda$ of $(X_0,Y_0)$, we have
\begin{equation*}
\boldsymbol{E}_{\lambda}\left[\|X_0-Y_0\|^2\right] \leq \rho^2_2(\nu_0,\mu_0)+\epsilon.
\end{equation*}
Since $\epsilon$ was arbitrary, we get
\begin{align*}
\Bigl|J^{\beta}(\nu_0)-J^{\beta}(\mu_0)\Bigr| \leq \left(\frac{\rho_2(\nu_0,\mu_0)}{1-\alpha}+\frac{2K_1}{1-\sqrt{\alpha}}\right)\rho_2(\nu_0,\mu_0),
		\end{align*}
where $K_1=\sqrt{K_2}$.
\end{proof} 

Given the equicontinuity result of Lemma~\ref{lem1Lin}, we have that
$J^{\beta}(\pi)$ is (uniformly)  continuous. Note that the proof of the lemma also
applies almost verbatim to the finite horizon case so that the bound in
Lemma~\ref{lem1Lin} also holds for the discounted finite horizon optimal
cost for all $T\ge 1$, which implies that the family of functions
$\{J^{\beta}(\pi,T): T\ge 1\}$ is (uniformly) equicontinuous on $\mathcal{Z}$. 

\begin{cor}\label{eqcontcor}
For any $T\ge 1$,  $\mu_0,\nu_0 \in {\cal Z}$, and  $\beta \in (0, 1)$, 
\begin{align}
  & \left|J^{\beta}(\nu_0,T)-J^{\beta}(\mu_0,T)\right| \nonumber \\
  &\quad  \leq  \left(\frac{\rho_2(\nu_0,\mu_0)}{1-\alpha}+\frac{2K_1}{1-\sqrt{\alpha}}\right)\rho_2(\nu_0,\mu_0).  \label{eq:eqcontbound1}
\end{align}
\end{cor}

\begin{proof}[Proof of Theorem~\ref{discountedThm}]
  
With Lemmas \ref{qccompact}--\ref{ctsKernel}  and Theorem
\ref{measurableselectionThm}, Condition \ref{condition1} (the measurable
selection condition) is satisfied  and the  function $v^*$ is lower semi-continuous, so
we can now define the so-called  value iteration (VI) updates recursively   (see,
e.g., \cite[(4.2.2)]{HeLa96}): For any $\pi \in {\cal Z}$, let 
\begin{equation}\label{VI}
v_n(\pi) = \min_{Q\in \mathcal{Q}_c} \left(c(\pi,Q)+\beta \int_{{\cal Z}} v_{n-1}(\pi')P(d\pi'|\pi,Q)\right),
\end{equation}
$ \enskip \text{for }n\geq 1$ with $v_0(\pi)=0$ for all $\pi$. Since $v_0 \equiv 0$ is continuous for
  $n=1$, we get that $v_1=\min_Q c(\pi,Q)$, and since $c(\pi,Q)$ is lower
  semi-continuous and ${\cal Q}_c$ is compact, we obtain that $v_1$ is also
  lower semi-continuous. For $n \geq 2$, by Theorem
  \ref{measurableselectionThm}, the iterations are well defined and
  $v_n$ is lower semi-continuous for all $n$.

It is known that $v_n$ is the value function of the $n$-stage
discounted cost $J^{\beta}(\pi,\Pi,n)$ in (\ref{Discfinite}) with zero terminal cost (see \cite[Chapter 4, p.45]{HeLa96}), i.e.,
\begin{align} \label{eq:vn}
	v_n(\pi) = \inf_{\Pi\in {\bf \Pi}_W^C} J^{\beta}(\pi,\Pi,n) =J^{\beta}(\pi,n) \quad \text{for all } \pi \in
  {\cal Z}. 
\end{align}

Note that, using the operator $\mathbb{H}$ defined in \eqref{eq:Hdef}, we may rewrite the DCOE
(\ref{DCOE}) and the VI functions in (\ref{VI}) as

\begin{equation}
     v=\mathbb{H} v, \quad \text{and } v_n=\mathbb{H}v_{n-1}  \text{\ for } n\geq 1,
\end{equation}
respectively.



 In addition, note that  since $c(\pi,Q)$ is
  non-negative,  $\mathbb{H}$ is monotone, i.e., for $u$ and $u'$ if
  $u \geq u'$ then $\mathbb{H}u \geq \mathbb{H}u'$. Therefore, since we start
  from $v_0=0$, then  $v_n$ is a non-decreasing sequence of lower
  semi-continuous functions.  By (\ref{Jbetaupperbound}) and (\ref{eq:vn})  we know that for all
  $n \geq 0$,
\begin{equation}
J^{\beta}(\pi) \geq  v_n(\pi). 
\nonumber 
\end{equation}
Thus $v_n$ is a non-decreasing and bounded sequence and hence it converges
pointwise to some function~$v $.  Since  $v_n(\pi) = J^{\beta}(\pi,n)$,  by
Corollary~\ref{eqcontcor}  $v_n$ is continuous and the sequence $\{v_n:
n\ge 1\}$  is a (uniformly) equicontinuous
which converges pointwise (on the metric space $\mathcal{Z}$). Therefore, the
limit function $v$ is continuous.

Now since both $v_n$ and $v$ are continuous we have that, by Lemma \ref{ctsc}
and Lemma \ref{ctsKernel}, the functions
\[ V^{\beta}_n(\pi, Q) \coloneqq \bigg(c(\pi,Q)+\beta \int_{{\cal Z}} v_n(\pi')P(d\pi'|\pi,Q)\bigg)\]
and
\[ V^{\beta}(\pi, Q) \coloneqq \bigg(c(\pi,Q)+\beta \int_{{\cal Z}}
  v(\pi')P(d\pi'|\pi,Q)\bigg)\] are continuous in $Q$, for each fixed $\pi$ for
all $n\ge 1$. Also, as $v_n \uparrow v$, by the dominated convergence theorem
$V^{\beta}_n(\pi, Q) \uparrow V^{\beta}(\pi, Q)$ for all
$(\pi,Q)\in \mathcal{Z}\times \mathcal{Q}_c$. Thus by \cite[Lemma 4.2.4]{HeLa96}, we can change the order of
limit and minimum as 
\[
 \lim_{n\to \infty} \min_{Q\in \mathcal{Q}_c}  V^{\beta}_n(\pi, Q) = \min_{Q\in
   \mathcal{Q}_c}   V^{\beta}_n(\pi, Q).
\]
Since the left hand side is $\lim_{n\to \infty} v_{n+1} =v$ and the right hand
side is $\mathbb{H} v$, we obtain the DCOE $v=\mathbb{H}v$, i.e., for all
$\pi\in \mathcal{Z}$, 
\begin{align}
 \label{eq:vpmsc}
v(\pi) &= \min_{Q\in \mathcal{Q}_c}  V^{\beta}(\pi, Q)  \nonumber \\
&= \min_{Q\in \mathcal{Q}_c}  \bigg(c(\pi,Q)+\beta \int_{{\cal Z}} v(\pi')P(d\pi'|\pi,Q)\bigg). 
\end{align}
According to Theorem~\ref{measurableselectionThm}, the measurable selection
condition Assumption~\ref{measurableSelectionCondition} holds in
\eqref{eq:vpmsc} (with $u^*=u$) and therefore there exists a (measurable) $\hat{\eta}:
\mathcal{Z}\to \mathcal{Q}_c$ such that for all
$\pi\in \mathcal{Z}$, 
\begin{equation}
 \label{eq:vpmsc1}
v(\pi) =    \bigg(c(\pi,\hat{\eta}(\pi))+\beta \int_{{\cal Z}} v(\pi')P(d\pi'|\pi,\hat{\eta}(\pi))\bigg). 
\end{equation}
Thus to finish the proof of the theorem we need only show that $v=J^{\beta}$ and
that the stationary and deterministic policy $\Pi=\{\hat{\eta}\}$ is
optimal. This is done with the aid of the following lemma which has a simple
proof (see, e.g., \cite[Lemma~5.4.4]{yuksel2020control}).
\begin{lemma}
  \label{lemver}
Assume  $v(\pi_0) =\lim_{n\to \infty} J^{\beta}(\pi_0,n)$  satisfies the DCOE
$v=\mathbb{H}v$ and the  stationary and deterministic policy
$\Pi=\{\hat{\eta}\}$ is such that it satisfies \eqref{eq:vpmsc1}. Assume
furthermore that
\begin{equation}
  \label{eq:vlim}
  \lim_{t \to \infty}  \beta^t \boldsymbol{E}^{\Pi}_{\pi_0}[v(\pi_t)]=0,
\end{equation}
for all $\pi_0\in \mathcal{Z}$, where $\{\pi_t\}$ is the state process of our MDP with initial distribution
$\pi_0$ and policy $\Pi$. Then $v(\pi_0)=J^{\beta}(\pi_0)$ and
$J^{\beta}(\pi_0,\Pi) = J^{\beta}(\pi_0)$ for all $\pi_0\in \mathcal{Z}$, i.e.,
$\Pi\in {\bf \Pi}_{WS}^C$ is an optimal policy. 
\end{lemma}

Note that  we have already shown that the first two conditions of the lemma
hold, so we have only
to check that \eqref{eq:vlim} holds in our case. By the bound \eqref{Jbetabound} in the proof of Lemma~\ref{JbetaLemma}, for any
initial condition $\pi_0$ and policy $\Pi$, 
we have 
\begin{align}\label{uniformbound}
  v(\pi_t) &\le   J^{\beta}(\pi_t) \le \frac{1}{1-\beta} \Big(
               \boldsymbol{E}_{\pi_t} [\|X\|^2] + \frac{1}{1-\alpha}
               \sigma^2\Big) \nonumber  \\
           & = \frac{1}{1-\beta} \Big(
               \boldsymbol{E}^{\Pi}_{\pi_0} \big[ \|X_t\|^2| q_{[0,\ldots,t-1]} \big]  +
             \frac{1}{1-\alpha} \sigma^2\Big).  \nonumber 
\end{align}
Thus
\begin{align*}
  \boldsymbol{E}^{\Pi}_{\pi_0}[v(\pi_t)]  & \le
  \frac{1}{1-\beta} \Big( \boldsymbol{E}^{\Pi}_{\pi_0} \big[ \|X_t\|^2 \big]  +
                                              \frac{1}{1-\alpha} \sigma^2\Big) \\
  &\le  \frac{1}{1-\beta} \Big( \boldsymbol{E}_{\pi_0} \big[ \|X_0\|^2 \big]  +
                                              \frac{2}{1-\alpha} \sigma^2\Big),
\end{align*}
where the last inequality holds by \eqref{eq3Moment}. Therefore, since $\beta \in (0,1)$,
\begin{equation*} 
\lim_{t \to \infty}  \beta^t \boldsymbol{E}^{\Pi}_{\pi_0}[v(\pi_t)]=0.
\end{equation*}

In summary, we have shown that $J^{\beta}$ satisfies the DCOE and
there exists a stationary deterministic policy $\Pi$ that is optimal. This finishes the proof of
Theorem~\ref{discountedThm}. 
\end{proof}

\subsection{Proof of  Theorem~\ref{thm1Lin}} \label{subsecAv}

This section is devoted to proving our main result. The
proof is done by  showing the existence of a so-called
canonical triplet for our MDP (see Definition~\ref{ACOEdef} in Appendix~A) which in term,
after checking that the conditions of 
Theorem~\ref{verificationACOE} in Appendix~A,  proves the existence of optimal stationary
and deterministic quantization policies and the stated convergence rate. 
Due to the nature of the controlled Markov process $\{\pi_t,Q_t\}$ in our
problem, verifying these sufficient conditions is technically challenging. 

In the following we present the proof of
Theorem \ref{thm1Lin}, our main result.
	
\begin{proof}[Proof of Theorem \ref{thm1Lin}]
		
  We will prove Theorem~\ref{thm1Lin} via the approach of vanishing
  discounted cost (see \cite[Chapter 5.3]{HeLa96}).  Recall that our state space
  is $\mathcal{Z} \coloneqq \mathcal{G} \cap \mathcal{P}_2(\mathbb{R}^d)$. In
  \cite{YukLinZeroDelay}, it was shown that, ${\cal G}$ is closed in
  $\mathcal{P}(\mathbb{R}^d)$. Note that
  $\mathcal{P}_2(\mathbb{R}^d) = \bigcup_{m \in \mathbb{N}} Z_m$, where
\begin{equation*}
Z_m =  \left\{ \mu \in \mathcal{P}_2(\mathbb{R}^d): \int_{\mathbb{R}^d} \|x\|^2 \mu(dx)  \leq m \right\},
\end{equation*}
and this implies that $\mathcal{Z}$ is $\sigma$-compact.

Next note that by Lemma
\ref{lem1Lin}, the family of functions
\begin{equation}
  \label{eq:hdef}
  h_{\beta}(\pi) \coloneqq J^{\beta}(\pi) - J^{\beta}(\mu), \quad \pi \in
\mathcal{Z}, 
\end{equation}
with fixed $\mu \in \mathcal{Z}$ is equicontinuous on $\mathcal{Z}$.
Theorem~\ref{discountedThm} in the previous section proved that $J^{\beta}$
solves the equation
\begin{equation}\label{DCOE1}
  J^{\beta}(\pi)= \min_{Q\in \mathcal{Q}_c}\left(c(\pi,Q) + \beta \int_{{\cal Z}} J^{\beta}(\pi')P(d\pi'|\pi,Q)\right).
\end{equation} 
With the definition of $h_{\beta}$ and an elementary calculation we can rewrite (\ref{DCOE1}) as
\begin{align}\label{DCOE2}
	&(1-\beta)J^{\beta}(\mu)+h_{\beta}(\pi) \nonumber \\
	& \quad =  \min_{Q\in \mathcal{Q}_c} \left(c(\pi,Q)+\beta\int_{\cal Z}h_{\beta}(\pi')P(d\pi'|\pi,Q)\right).
\end{align}
		
Recall that by Lemma \ref{JbetaLemma} for all $\beta \in (0,1)$, we have the upper bound
\begin{align*}
(1-\beta)J^{\beta}(\mu) \leq	\boldsymbol{E}_{\mu}[\|X\|^2] + \frac{1}{1-\alpha} \sigma^2,  
\end{align*}
which is independent of $\beta$. 

Since the range of $(1-\beta)J^{\beta}(\mu)$, $\beta\in (0,1)$  is bounded,  there exists a
limit point $\rho^*$  as $\beta \uparrow 1$. Let $\beta(l)$ be a sequence such that 
\begin{equation*}
\lim_{l \to \infty}  (1-\beta(l))J^{\beta(l)}(\mu) = \rho^*.
\end{equation*}
(Note that $\rho^*$ depends on the fixed $\mu\in \mathcal{Z}$, but not on
$\pi$.) 		
 By the conditions on the state space ${\cal Z}$, the equicontinuity of $h_{\beta}$, and the Arzela-Ascoli theorem (see Theorem~\ref{Ascoli} in Appendix~A), there exists a subsequence $\{h_{\beta(k)}\}$ of $\{h_{\beta(l)}\}$ which converges pointwise to a continuous function $h$ 
 \begin{equation}\label{hlim}
 h(\pi)\coloneqq\lim_{k \rightarrow \infty} h_{\beta(k)}(\pi), \quad \pi \in {\cal Z}.
 \end{equation}
 Then, (\ref{DCOE2}) along the subsequence $\beta(k)$ becomes
 \begin{align}\label{DCOE3}
	&(1-\beta(k))J^{\beta(k)}(\mu)+h_{\beta(k)}(\pi)  \nonumber \\
	&=  \min_{Q\in \mathcal{Q}_c} \left(c(\pi,Q)+\beta(k)\int_{\cal Z}h_{\beta(k)}(\pi')P(d\pi'|\pi,Q)\right).
\end{align} 
If we take the limit of (\ref{DCOE3}) as $k \to \infty$ we get
 \begin{align}\label{ACOELim}
	&\rho^*+h(\pi) \nonumber \\
	&=\lim_{k \rightarrow \infty}  \min_{Q\in \mathcal{Q}_c} \left(c(\pi,Q)+\beta(k)\int_{\cal Z}h_{\beta(k)}(\pi')P(d\pi'|\pi,Q)\right). 
\end{align} 
	
Define
\begin{align*}
  V_{k}(\pi,Q) &\coloneqq c(\pi,Q)+\beta(k)\int_{\cal Z}h_{\beta(k)}(\pi')P(d\pi'|\pi,Q), \quad  \\
  V(\pi,Q) &\coloneqq c(\pi,Q)+\int_{\cal Z}h(\pi')P(d\pi'|\pi,Q).
\end{align*}
In the following we show that average cost optimality equation (ACOE) in
Definition~\ref{ACOEdef} in Appendix~A holds, i.e.,
\begin{equation*}
\rho^*+h(\pi) = \min_{Q\in \mathcal{Q}_c} \left(c(\pi,Q)+\int_{\cal
    Z}h(\pi')P(d\pi'|\pi,Q)\right),
\end{equation*}
        $ \text{ \ for all $\pi\in \mathcal{Z}$}$.
\begin{lemma}\label{limmin1}
      	Consider $V_k(\pi,Q)$ and $V(\pi,Q)$ defined above. Then, 
        	$$\lim_{k \rightarrow \infty}  \min_{Q\in \mathcal{Q}_c} V_k(\pi,Q) =  \min_{Q\in \mathcal{Q}_c} V(\pi,Q).$$
\end{lemma}
\begin{proof}
  Let $Q_k^*$ and $Q^*$ minimize $V_k(\pi,Q)$ and $V(\pi,Q)$ respectively. Then
  by the definition of $V_k$ and $V$, it suffices to show that the upper bound
  in the following inequality converges to zero as $k \to \infty$:

 \begin{eqnarray}
  &&\bigg| \min_{Q\in \mathcal{Q}_c} \Big ( c(\pi,Q) + \int_{\cal
   Z}\beta(k) h_{\beta(k)}(\pi')P(d\pi'|\pi,Q)\Big)  \nonumber \\
  && \quad  -  \min_{Q\in
   \mathcal{Q}_c} \Big( c(\pi,Q)+ \int_{\cal
   Z}   h(\pi')P(d\pi'|\pi,Q)\Big)  \bigg|   \nonumber\\  
  &&\leq \max \Big( \Big| \int_{\cal
     Z} \beta(k)  h_{\beta(k)}(\pi')P(d\pi'|\pi,Q^*_k) \nonumber \\
     &&  \quad \quad \quad \quad \quad \quad - \int_{\cal
        Z}h(\pi')P(d\pi'|\pi,Q^*_k)\Big|, \label{1stterm} \\ 
        & & \qquad \Big|\int_{\cal Z} \beta(k) h_{\beta(k)}(\pi')P(d\pi'|\pi,Q^*)  \nonumber \\
     &&  \quad \quad \quad \quad\quad \quad - \int_{\cal Z}h(\pi')P(d\pi'|\pi,Q^*)\Big|\Big), \label{max}
 \end{eqnarray}
 Since $\beta(k)\uparrow 1$, it is enough to show that \eqref{1stterm} and
 \eqref{max} go to zero as $k\to \infty$ with the $\beta(k)$
 multiplicative terms replaced by $1$. 

Let 
\begin{equation}\label{eqg}
	  g(\pi)\coloneqq\left(\frac{\rho_2(\pi,\mu)}{1-\alpha}+\frac{2K_1}{1-\sqrt{\alpha}}\right)\rho_2(\pi,\mu),  
\end{equation}
where $\mu \in {\cal Z}$ is given from the definiton of $h_{\beta}$ in \eqref{eq:hdef}.
Observe that by Lemma~\ref{lem1Lin},
\begin{equation}\label{eqgUpper}
    		|h_{\beta(k)}(\pi)| \leq g(\pi) < \infty,
\end{equation}
for all $\pi \in {\cal Z}$. 
Note that by choosing the joint measure so that the marginals are independent, the Wasserstein distance $\rho_2$ can be upper bounded as
\begin{equation}\label{ro2upper}
  \rho^2_2(\pi,\mu) \leq 2\boldsymbol{E}_{\pi}[\|X\|^2]+2\boldsymbol{E}_{\mu}[\|X\|^2]< \infty.  
\end{equation}
    	 
Now we show that the term in (\ref{1stterm}) converges to zero (the convergence
of \eqref{1stterm} will follow from this proof too). Suppose
otherwise that for some $\epsilon > 0$ there exists a subsequence $Q^*_{k_l}$
such that
\begin{equation}\label{contradict}
    	 \Big|\int_{\cal Z}h_{\beta(k_{l})}(\pi')P(d\pi'|\pi,Q^*_{k_l}) - \int_{\cal Z}h(\pi')P(d\pi'|\pi,Q^*_{k_l})\Big| \geq \epsilon.
\end{equation}
By the compactness of ${\cal Q}_c$ there exists further subsequence
$Q^*_{k_{l_n}}$ that converges to a quantizer $\bar{Q} \in {\cal Q}_c$. In the
following we prove that along the subsequence $Q^*_{k_{l_n}}$ the term on the
left hand side of (\ref{contradict}) goes to zero and reach a contradiction.
To do so, we use Lemma \ref{langen} in the Appendix. Note that by
(\ref{eqgUpper}) and (\ref{ro2upper}), for all $n \geq 1$ we have
 \begin{align}
    	 	|h_{\beta(k_{l_n})}(\pi)| \leq g(\pi) \leq
   g_1(\pi), \label{eq:hbupp} 
 \end{align}
 where
 \begin{align}
  & g_1(\pi) \coloneqq \frac{2}{1 -
   \alpha}\left(\boldsymbol{E}_{\pi}[\|X\|^2]+\boldsymbol{E}_{\mu}[\|X\|^2]\right) \nonumber \\
& \qquad \qquad   + \frac{2K_1}{1-
   \sqrt{\alpha}}\sqrt{2\boldsymbol{E}_{\pi}[\|X\|^2]+2\boldsymbol{E}_{\mu}[\|X\|^2]}.
     \label{eq:hbupp1}   
 \end{align}
 Furthermore, for any sequence $\{\pi_n\} \in {\cal Z}$, with
 $\rho_2(\pi_n,\pi) \to 0$, we have that $h_{\beta(k_{l_n})}$ continuously
 converges to $h$ (i.e.
 $\lim_{n \to \infty} h_{\beta(k_{l_n})}(\pi_n) = h(\pi)$) since
 \begin{align*}
    	 & |h_{\beta(k_{l_n})} (\pi_n) - h(\pi)| \\
	 &\quad \leq | h_{\beta(k_{l_n})} (\pi_n) - h_{\beta(k_{l_n})} (\pi)| + | h_{\beta(k_{l_n})} (\pi) - h(\pi)| \\
	  &\quad = | J^{\beta(k_{l_n})} (\pi_n) - J^{\beta(k_{l_n})} (\pi)| + |h_{\beta(k_{l_n})} (\pi)- h(\pi)| \\  
	  &\quad \leq \left(\frac{\rho_2(\pi_n,\pi)}{1-\alpha}+\frac{2K_1}{1-\sqrt{\alpha}}\right)\rho_2(\pi_n,\pi) \\
&\qquad \qquad	  +  |h_{\beta(k_{l_n})} (\pi)- h(\pi)|  \to 0, \quad \text{as } n \to \infty,
 \end{align*}
 where the last inequality follows from Lemma \ref{lem1Lin}. Since $Q^*_{k_{l_n}} \to \bar{Q}$ in order-2 Wasserstein distance and since
$g_1$ is continuous for the order-2 Wasserstein convergence of its argument, we
have by Lemma~\ref{ctsKernel}, 
 \begin{align*}
       	 	\lim_{n \to \infty} \int_{{\cal Z}} g_1(\pi') P(d\pi'|\pi,Q^*_{k_{l_n}}) = \int_{{\cal Z}} g_1(\pi') P(d\pi'|\pi,\bar{Q}) < \infty.
\end{align*} 
Hence the conditions of
the generalized dominated  convergence theorem in Lemma~\ref{langen} in Appendix~A
are satisfied, which gives 
\[
	\lim_{n \to \infty} \int_{{\cal Z}} h_{\beta(k_{l_n})}(\pi')  P(d\pi'|\pi,Q^*_{k_{l_n}}) = \int_{{\cal Z}} h(\pi') P(d\pi'|\pi,\bar{Q}).
 \]
Since we also clearly have
\[
	\lim_{n \to \infty} \int_{{\cal Z}} h(\pi')  P(d\pi'|\pi,Q^*_{k_{l_n}}) = \int_{{\cal Z}} h(\pi') P(d\pi'|\pi,\bar{Q}),
\] 
we obtain 
 \begin{align*}
    	 	\int_{\cal Z}h_{\beta(k_{l_n})}(\pi')P(d\pi'|\pi,Q^*_{k_{l_n}}) - \int_{\cal Z}h(\pi')P(d\pi'|\pi,Q^*_{k_{l_n}}) \to 0,
\end{align*}
which contradicts (\ref{contradict}). Hence the term in (\ref{1stterm}) also
goes to zero and this concludes the proof.
 \end{proof}
  Thus, by Lemma \ref{limmin1} we can change the order of limit and minimum in
  (\ref{ACOELim}), then we get 
 \begin{align*}
			& \rho^*+h(\pi) \\
			&=  \min_{Q\in \mathcal{Q}_c} \lim_{k \rightarrow \infty} \left(c(\pi,Q)+\beta(k)\int_{\cal Z}h_{\beta(k)}(\pi')P(d\pi'|\pi,Q)\right) \\
			&=  \min_{Q\in \mathcal{Q}_c} \left(c(\pi,Q)+\int_{\cal Z}h(\pi')P(d\pi'|\pi,Q)\right) \\
			&= c(\pi,Q^*)+\int_{\cal Z}h(\pi')P(d\pi'|\pi,Q^*).
 \end{align*}
 Noting that $Q^*=Q^*_{\pi}$ is a function of $\pi$ in the last equation and
 defining $\hat{\eta}: \mathcal{Z}\to \mathcal{Q}_c$ by $\hat{\eta}(\pi)=
 Q^*_{\pi}$, we obtain that ($\rho^*,h,\hat{\eta}$) is a canonical triplet for
 which the ACOE holds (see Definition~\ref{ACOEdef} in Appendix~A). 

 Now we are ready to apply Theorem~\ref{verificationACOE} in Appendix~A  to 
 complete the proof of Theorem~\ref{thm1Lin}. For this recall that  for all  $\pi \in \mathcal{Z}$, by \eqref{eq:hbupp} and
 \eqref{eq:hbupp1} we have 
 \begin{align}
|h_{\beta}(\pi)|    &=|J^{\beta}(\pi)-J^{\beta}(\mu)| \nonumber \\
 &\leq \frac{2}{1 - \alpha}\left(\boldsymbol{E}_{\pi}[\|X\|^2]+\boldsymbol{E}_{\mu}[\|X\|^2]\right)  \nonumber \\
& \quad + \frac{2K_1}{1- \sqrt{\alpha}}\sqrt{2\boldsymbol{E}_{\pi}[\|X\|^2]+2\boldsymbol{E}_{\mu}[\|X\|^2]}. \label{Kupper1}
\end{align}
Fix the initial distribution $\pi_0$, let $\Pi\in {\bf \Pi}_W^C$ be arbitrary, and let
$\{\pi_t\}$ be the states generated by this policy. Since the inequality in
(\ref{Kupper1}) holds for all $\pi \in {\cal Z}$, in particular it holds for
$\pi_T \in {\cal Z}$. Thus, from (\ref{hlim}) and (\ref{Kupper1}) we get
\begin{align}
& |h(\pi_T)| =\lim_{k \to \infty} |h_{\beta(k)}(\pi_T)|  \nonumber\\ 
&\leq  \frac{2}{1 - \alpha}\bigg(\boldsymbol{E}_{\pi_T}[\|X\|^2] \nonumber\\ 
& \quad +\boldsymbol{E}_{\mu}[\|X\|^2]\bigg) + \frac{2K_1}{1- \sqrt{\alpha}}\sqrt{2\boldsymbol{E}_{\pi_T}[\|X\|^2]+2\boldsymbol{E}_{\mu}[\|X\|^2]}. \label{hpiTupper}
\end{align}
Note that 
 \begin{align}
&	\boldsymbol{E}_{\pi_0} \left[\boldsymbol{E}_{\pi_T}[\|X\|^2] \right]  \nonumber \\*
	&= \boldsymbol{E}_{\pi_0} \left[\boldsymbol{E}[\|X_T\|^2|q_{[0,\ldots,T-1]} \right] = \boldsymbol{E}_{\pi_0} [\|X_T\|^2]  \nonumber\\ &\leq \boldsymbol{E}_{\pi_0} [\|X_0\|^2] + \frac{1}{1-\alpha}\sigma^2, \label{eqpiTUpper}
 \end{align}
where the inequality follows from \eqref{eq3Moment}. 
Now choose $\mu$ in the definition of $h_{\beta}$ as $\mu=\pi_0$. Then  (\ref{eqpiTUpper}),  (\ref{hpiTupper}),  and Jensen's inequality
give
\begin{align}
&		\boldsymbol{E}^{\Pi}_{\pi_0} [|h(\pi_T)|] \leq \frac{2}{1 -
  \alpha}\left(2\boldsymbol{E}_{\pi_0}[\|X\|^2] +
  \frac{1}{1-\alpha}\sigma^2\right) \nonumber \\
&  \quad \quad + \frac{2K_1}{1-
  \sqrt{\alpha}}\sqrt{4\boldsymbol{E}_{\pi_0}[\|X\|^2]+
  \frac{2}{1-\alpha}\sigma^2}. \label{eq:hT}
\end{align}
Hence, we have
 \begin{align*}
	 	\limsup_{T \to \infty}\frac{1}{T}
   \boldsymbol{E}^{\Pi}_{\pi_0}\left[h(\pi_T)\right]=0,
 \end{align*}
 for all $\pi_0$ and under every policy $\Pi$. Therefore by
 Theorem~\ref{verificationACOE}  there exists a
 deterministic stationary policy $\Pi^* \in {\bf \Pi}^C_{WS}$ that achieves the
 minimum in (\ref{mainthm}) simultaneously for all $\pi_0$.  Furthermore, by
 Theorem \ref{verificationACOE}
 \begin{align*}
&	 \left|J(\pi_0,\Pi^*,T)-J(\pi_0,\Pi^*)\right|  \\
&\qquad	 \leq \frac{1}{T}\left(\left|\boldsymbol{E}_{\pi_0}^{\Pi^*}[h(\pi_T)] - h(\pi_0)\right|\right) \\
&\qquad	 = \frac{1}{T}\left(\left|\boldsymbol{E}_{\pi_0}^{\Pi^*}[h(\pi_T)]\right|\right) \leq \frac{K(\pi_0)}{T},
 \end{align*}
 where the equality follows by the definition of $h$, 
 \begin{align*}
	 	h(\pi_0) = \lim_{k \to \infty} h_{\beta(k)}(\pi_0) = \lim_{k \to \infty} (J^{\beta(k)}(\pi_0) - J^{\beta(k)}(\pi_0)) = 0,
 \end{align*}
and  $K(\pi_0)$ is the upper bound in \eqref{eq:hT}. 
This concludes the the proof of the  second part of  Theorem \ref{thm1Lin}.
\end{proof}

\section{Conclusion}
In this paper we have considered the problem of zero-delay coding of
$\mathbb{R}^d$-valued linearly generated Markov sources. We have proved
structural, existence, and converge rate results for  optimal  zero-delay
coding under the assumption that the allowable quantizers have convex
codecells. Applications  to closed-loop control systems,
especially to optimal quadratic control under information constraints for
infinite horizons, see e.g.,
\cite{BaoMikaelKalle}--\cite{yuksel2019note},
are currently under study.
	
\section{Appendix A}

\subsection{Average Cost Optimality in Markov Decision Processes}
Let $\mathcal{Z}$ be a Borel space (i.e., a Borel subset of a complete and
separable metric space) and let $\mathcal{P}(\mathcal{Z})$ denote the set of all
probability measures on $\mathcal{Z}$. A discrete time Markov control model
(Markov decision process) is a system characterized by the 4-tuple
$ (\mathcal{Z}, \mathcal{A}, \mathcal{K}, c),$ where (i) $\mathcal{Z}$ is the
state space, the set of all possible states of the system; (ii) $\mathcal{A}$ (a
Borel space) is the control space (or action space), the set of all controls
(actions) $a \in \mathcal{A}$ that can act on the system; (iii)
$\mathcal{K} = \mathcal{K}(\,\cdot\,|z,a)$ is the transition probability of the
system, a stochastic kernel on $\mathcal{Z}$ given
$\mathcal{Z} \times \mathcal{A}$, i.e., $\mathcal{K}(\, \cdot\,|z, a)$ is a
probability measure on $\mathcal{Z}$ for all state-action pairs $(z, a)$, and
$\mathcal{K}(B| \, \cdot\, , \, \cdot\, )$ is a measurable function from
$\mathcal{Z} \times \mathcal{A}$ to $[0, 1]$ for each Borel set
$B \subset \mathcal{Z}$; (iv)
$c : \mathcal{Z} \times \mathcal{A} \rightarrow [0,\infty)$ is the cost per time
stage function of the system, a Borel measurable function $c(z,a)$ of the state
and the control.

Define the \emph{history} spaces $\mathcal{H}_t$ at time $t\ge 0$ of the Markov control
model by  $\mathcal{H}_0 \coloneqq \mathcal{Z}$ and $\mathcal{H}_t \coloneqq (\mathcal{Z}\times \mathcal{A})^t  
 \times \mathcal{Z}$.  Thus a specific history  $h_t \in \mathcal{H}_t$ has
the form $h_t = (z_0,a_0,\ldots,z_{t-1},a_{t-1},z_t)$.

\begin{definition}[Admissible Control Policy \cite{HeLa96}]
  An \emph{admissible control policy} $\Pi = \{\alpha_t\}_{t \geq 0}$, also
  called a \emph{randomized control policy} (more simply a \emph{control policy}
  or a \emph{policy}) is a sequence of stochastic kernels on the action space
  $\mathcal{A}$ given the history $\mathcal{H}_t$.  The set of all randomized control policies is
  denoted by ${\bf \Pi}_A$. A \emph{deterministic policy} $\Pi$ is a sequence of
  functions $\{\alpha_t\}_{t \geq 0}$, $\alpha_t : \mathcal{H}_t \to \mathcal{A}$, that
  determine the control used at each time stage deterministically, i.e., $a_t =
  \alpha_t(h_t)$. The set of all deterministic policies is denoted
  $\Pi_D$.  Note that $\Pi_D \subset {\bf \Pi}_A$. A \emph{Markov policy} is a policy
  $\Pi$ such that for each time stage the choice of control only depends on the
  current state $z_t$, i.e.,\ $\Pi = \{\alpha_t\}_{t \geq 0}$ with  $\alpha_t
  : \mathcal{Z} \to \mathcal{P}(\mathcal{A})$.  The set of all Markov policies is denoted by
  $\Pi_M$. The set of deterministic Markov policies is denoted by
  $\Pi_{MD}$. A \emph{stationary policy} is a Markov policy $\Pi =
  \{\alpha_t\}_{t \geq 0}$ such that $\alpha_t = \alpha$ for all $t \geq 0$ for some
   $\alpha : \mathcal{Z} \to \mathcal{P}(\mathcal{A})$.  The set of all stationary policies
  is denoted by $\Pi_S$ and the set of deterministic stationary policies is denoted
  by $\Pi_{SD}$. 
\end{definition}

The transition kernel $\mathcal{K}$, an initial probability distribution $\pi_0$ on
$\mathcal{Z}$, and a policy $\Pi$ define a unique probability measure
$P_{\pi_0}^{\Pi}$ on
$\mathcal{H}_{\infty}=(\mathcal{Z}\times\mathcal{A})^{\infty}$, the distribution
of the state-action process $\{(Z_t,A_t)\}_{t\ge 0}$. The resulting state
process $\{Z_t\}_{t\ge 0}$ is called a \emph{controlled Markov process}.  The
expectation with respect to $P_{\pi_0}^{\Pi}$ is denoted by
$\boldsymbol{E}_{\pi_0}^{\Pi}$. If $\pi_0=\delta_z$, the point mass at $z\in \mathcal{Z} $, we write $P_{z}^{\Pi}$ and $\boldsymbol{E}_{z}^{\Pi}$ instead of
$P_{\delta_z}^{\Pi}$ and $\boldsymbol{E}_{\delta_z}^{\Pi}$.
In an \emph{optimal control problem}, a performance objective $J$ of the system
is given and the goal is to find the controls that minimize (or maximize) that
objective.  Some common optimal control problems for Markov control models are
the following:

\begin{enumerate}
\item \emph{Finite Horizon Average Cost Problem}: Here the goal is to find policies that minimize the average
cost
\[
J(\pi_0,\Pi,T) \coloneqq \boldsymbol{E}^{\Pi}_{\pi_0}\left[\frac{1}{T} \sum_{t=0}^{T-1} c(Z_t,A_t)\right],
\]
for some $T \ge 1$.

\item \emph{Infinite Horizon Discounted Cost Problem}:
Here the goal is to find policies that minimize
\[
J^{\beta}(\pi_0,\Pi) \coloneqq \lim_{T \to \infty} \boldsymbol{E}^{\Pi}_{\pi_0}\left[ \sum_{t=0}^{T-1} \beta^t c(Z_t,A_t)\right],
\]
for some $\beta \in (0,1)$.

\item \emph{Infinite Horizon Average Cost Problem}:
In the more challenging infinite horizon control problem  the goal is to find
policies that minimize the average cost 
\[
J(\pi_0,\Pi) \coloneqq \limsup_{T \to \infty}
\boldsymbol{E}^{\Pi}_{\pi_0}\left[\frac{1}{T}   \sum_{t=0}^{T-1}
  c(Z_t,A_t)\right]. \label{infiniteCost1} 
\]
\end{enumerate}

The Markov control model together with the performance objective is called a
\emph{Markov decision process} (MDP).

\begin{definition}\cite{survey}\label{ACOEdef} Let $h$ and $g$ be measurable real functions on $\mathcal{Z}$
and let $f : \mathcal{Z} \rightarrow \mathcal{A}$ be measurable. Then $(g, h, f )$ is said to be a canonical triplet if for all $z \in \mathcal{Z}$,
\begin{equation}\label{can1}
g(z)=\inf_{a \in \mathcal{A}} \int_{\mathcal{Z}}g(z')\mathcal{K}(dz'|z,a) 
\end{equation}
\begin{equation}\label{can2}
g(z) + h(z)=\inf_{a \in \mathcal{A}} \left(c(z,a)+\int_{\mathcal{Z}}h(z')\mathcal{K}(dz'|z,a)\right)
\end{equation} 
and
\begin{equation}\label{can3}
g(z)= \int_{\mathcal{Z}}g(z')\mathcal{K}(dz'|z,f(z))
\end{equation}
\begin{equation}\label{can4}
g(z) + h(z)= c(z,f(z))+\int_{\mathcal{Z}}h(z')\mathcal{K}(dz'|z,f(z)).
\end{equation}
Equations \eqref{can1}-\eqref{can2} and \eqref{can3}-\eqref{can4} are called the canonical equations.
In case $g$ is a constant, $g \equiv g^* $, these equations reduce to
\begin{equation}\label{ACOE}
g^* + h(z)=\inf_{a \in \mathcal{A}} \left(c(z,a)+\int_{\mathcal{Z}}g(z')\mathcal{K}(dz'|z,a)\right)
\end{equation}
\begin{equation}\label{can6}
g^* + h(z)= c(z,f(z))+\int_{\mathcal{Z}}h(z')\mathcal{K}(dz'|z,f(z)),
\end{equation}
and \eqref{ACOE}-\eqref{can6} is called the \textit{average cost optimality equation} (ACOE).
\end{definition}
\begin{theorem}\label{verificationACOE}\cite[Theorem 7.1.1]{yuksel2020control}
Let $(g,h,f)$  be a canonical triplet. If $g \equiv g^*$ is a constant and
\begin{equation*}
\limsup_{T \to \infty} \frac{1}{T}\boldsymbol{E}^{\Pi}_{z}[h(z_T)]=0,
\end{equation*}
for all $z$ and under every policy $\Pi\in {\bf \Pi}_A$, then the stationary
deterministic policy $\Pi^* = \{f\}\in \Pi_{SD}$ is optimal so that
\begin{equation*}
g^*=J(z,\Pi^*)=\inf_{\Pi\in {\bf \Pi}_A} J(z,\Pi),
\end{equation*}
where
\begin{equation*}
J(z,\Pi) = \limsup_{T \to \infty} \frac{1}{T} \boldsymbol{E}^{\Pi}_{z}\left[\sum_{t=0}^{T-1}c(z_t,a_t)\right] .
\end{equation*}
Furthermore,
\begin{equation*}
\left| \frac{1}{T} \boldsymbol{E}_{z}^{\Pi^*} \sum_{t=0}^{T-1}c(z_{t},a_{t}) - g^* \right| \leq \frac{1}{T} \left(\left|\boldsymbol{E}_z^{\Pi^*}[h(z_T)] - h(z)\right|\right),
\end{equation*} 
i.e. 
\begin{eqnarray}\label{eq38}
	\left|J(z,\Pi^*,T)-g^*\right|&=&
                                         \left|J(z,\Pi^*,T)-J(z,\Pi^*)\right|
                                         \nonumber \\
  &\leq & \frac{1}{T}\left(\left|\boldsymbol{E}_z^{\Pi^*}[h(z_T)] - h(z)\right|\right).
\end{eqnarray}
\end{theorem}

\subsection{Auxiliary Results}

\begin{theorem} \cite[Theorem 2.4.7]{Dud02} \label{Ascoli}
Let $F$ be an equicontinuous family of real functions on a compact space $\mathcal{X}$ and let $f_n$ be a sequence in $F$ such that the range of $f_n$ is compact. Then, there exists a subsequence $f_{n_k}$ which converges uniformly to a continuous function. If  $\mathcal{X}$ is $\sigma$-compact, $f_{n_k}$ converges pointwise to a continuous function, and the convergence is uniform on compact subsets of $\mathcal{X}$.
\end{theorem}

\begin{lemma}\label{mincts}
Let ${\cal A}$ be compact, and let $V(z,a)$ be continuous on ${\cal Z} \times {\cal A}$. Then, $\min_{a\in {\cal A}}V(z,a)$ is continuous on ${\cal Z}$.
\end{lemma}

\begin{lemma}\cite[Lemma 2]{YukLinZeroDelay}\label{Lem2LY2014}
\begin{itemize}
\item[\em (a)] Let $\{\pi_n\}$ be a sequence of probability density functions on
$\mathbb{R}^d$  which are  uniformly equicontinuous and uniformly
bounded  and assume $\pi_n\to \pi$ weakly. Then $\pi_n\to \pi$  in
total variation.
			
\item[\em (b)] Let $\{Q_n\}$ be a sequence in $\mathcal{Q}_c$ such that $Q_n \to
Q$ weakly at $P$ for some $Q\in \mathcal{Q}_c$. If $P$ admits a density, then
$Q_n \to Q$ in total variation at $P$. If the density of  $P$ is positive, then
$Q_n \to Q$ in total variation at \emph{any} $P'$ admitting a density.
\end{itemize}
\end{lemma}

\begin{lemma}\cite[Theorem 3.5]{langen1981convergence}\label{langen}
Suppose $f_n, b_n, f,$ and $b$ are measurable real functions on a standard Borel
space ${\cal X}$. Let $\{\mu_n\}$ be a sequence of probability measures in
${\cal P}({\cal X})$, converging weakly to some $\mu \in {\cal P}({\cal
  X})$. Assume that  
\begin{align*}
|f_n| &\leq b_n, \quad n\geq 1,
\end{align*}
and that 
\begin{align*}
	f_n \xrightarrow{\text{c}} f, &\quad  b_n \xrightarrow{\text{c}} b, \\
	\lim_{n \to \infty} \int_{{\cal X}} b_n(x) \mu_n(dx)& = \int_{{\cal X}} b(x) \mu(dx) < \infty,
\end{align*}
where $f_n \xrightarrow{\text{c}} f$ means that for any $x \in {\cal X}$ and any
sequence $x_n \to x$, we have $f_n(x_n) \to f(x)$ (i.e., $f_n$ continuously converges to f).
Then,
\begin{align*}
\lim_{n \to \infty} \int_{{\cal X}} f_n(x) \mu_n(dx) = \int_{{\cal X}} f(x) \mu(dx).
\end{align*}
\end{lemma}
	
\section{Appendix B}\label{appB}
\begin{proof}[Proof of Lemma \ref{qccompact}]
  \cite[Lemma 3]{YukLinZeroDelay} shows that $\mathcal{Q}_c$ is compact in the weak
  topology, so by \cite[Theorem 7.12]{villani2003topics} we only need to prove the
  convergence of second moment, i.e., we have to show that for any
  $P \in {\cal Z}$,
\begin{equation} \label{eq:cconv}
\int_{\mathbb{R}^d\times {\cal M}} \|z\|^2dPQ_n \rightarrow \int_{\mathbb{R}^d\times {\cal M}} \|z\|^2dPQ,
\end{equation}
whenever $PQ_n \rightarrow PQ$ weakly.
Note that
\begin{equation*}
\int_{\mathbb{R}^d\times {\cal M}}\|z\|^2dPQ_n = \sum_{i=1}^{M} \int_{B_i^n}(\|x\|^2+i^2) P(dx),
\end{equation*}
where the $\{B_i^n\}$ are the bins of $Q_n$. 
		
For $\epsilon > 0$, let $L > 0$ be such  that
$\int_{\{\|x\| \geq L\}} \|x\|^2 P(dx) < \epsilon$. Letting $\{B_i\}$ be the
bins of $Q$, we have
\begin{align}
	&\left\| \int \|z\|^2dPQ_n - \int \|z\|^2dPQ \right\|\nonumber \\*
	&\quad \leq  \sum_{i=1}^{M} \int_{\mathbb{R}^d}(\|x\|^2+i^2) (|1_{B_i^n}-1_{B_i}|) P(dx) \nonumber \\ 
	&\qquad \leq \sum_{i=1}^{M} \Bigl(\int_{\mathbb{R}^d}\|x\|^2 (|1_{B_i^n}-1_{B_i}|) P(dx)\nonumber \\
	& \qquad + \int_{\mathbb{R}^d} M^2 (|1_{B_i^n}-1_{B_i}|) P(dx)\Bigr)\nonumber\\
	&\qquad = \sum_{i=1}^{M} \Biggl(\int_{\{\|x\|\geq L\}}\|x\|^2(|1_{B_i^n}-1_{B_i}|) P(dx)\nonumber \\
	&\qquad + \int_{\{\|x\| < L\}}\|x\|^2(|1_{B_i^n}-1_{B_i}|) P(dx) \Biggr)\nonumber\\ 
	&\qquad \qquad \qquad \qquad \qquad \qquad + M^2 \sum_{i=1}^{M} P(B_i^n \Delta B_i) \nonumber \\	
	&\qquad \leq \sum_{i=1}^{M} \Bigl(\epsilon + L^2 P(B_i^n \Delta
   B_i)\Bigr) + M^2 \sum_{i=1}^{M} P(B_i^n \Delta B_i) \nonumber  \\
	&\qquad \leq \epsilon M + (L^2+M^2) \sum_{i=1}^{M} P(B_i^n \Delta B_i) \rightarrow \epsilon M, \nonumber
\end{align}
as $n \rightarrow \infty$. It was shown in the proof of \cite[Lemma 2]{YukLinZeroDelay} that if
$PQ_n \rightarrow PQ$ weakly then by the assumption that $P$ admits a density we
have $P(B_i^n \Delta B_i) \rightarrow 0$ for all $i=1,\ldots,M$.
Since $\epsilon$ was arbitrary we obtain \eqref{eq:cconv}, which completes the
proof. 
\end{proof}

The next lemma is needed in the proof of  Lemma~\ref{ctsc}.
              
\begin{lemma}\label{receiverLemma}
  Let $\{B_1^n,\ldots,B_M^n\}$ and $\{B_1,\ldots,B_M\}$ denote the cells of quantizers
  $Q_n$ and $Q$ respectively. If  $(\pi_n,Q_n) \rightarrow (\pi,Q)$ in
  $\mathcal{Z}\times \mathcal{Q}_c$, the optimal receiver $\gamma_n$ for $Q_n$
  converges to optimal receiver $\gamma$ of $Q$ in the sense that
\[
\gamma_n(i) = \frac{1}{\pi_n(B^n_i)}\int_{B^n_i} x \, \pi_n(dx) \rightarrow
\frac{1}{\pi(B_i)}\int_{B_i} x\,  \pi(dx) =\gamma(i),
\]
for every $i \in \{1,\ldots,M\}$ such that $\pi(B_i)>0$.

\end{lemma}

\begin{proof} We have
  \begin{eqnarray}
  && \lefteqn{ \Big\| \int_{B^n_i} x \, \pi_n(dx) -
   \int_{B_i} x\,  \pi(dx)  \Big\|    } \nonumber \\*
   && \qquad \le  \Big\|  \int_{B^n_i} x \, \pi_n(dx) -
          \int_{B^n_i} x\,  \pi(dx)  \Big\| \nonumber \\
 && \qquad \qquad\quad         +
       \Big\| \int_{B^n_i} x \, \pi(dx) -   \int_{B_i} x\,  \pi(dx)  \Big\| 
          \nonumber \\
     && \qquad\le   \Big\| \int_{B^n_i} x  \, \pi_n(dx) -  \int_{B^n_i}x \,  \pi(dx)
            \Big\| \nonumber \\
            &&  \qquad \qquad \quad \quad +
        \int_{B^n_i\triangle B_i} \|x\| \, \pi(dx).           \label{eqabsc}
\end{eqnarray}
Since $(\pi_n,Q_n) \rightarrow (\pi,Q)$, we have 
$\pi(B_i^n\Delta B_i)\to 0$ (see \cite{YukselOptimizationofChannels}). Since
$E_{\pi}[ \|X\|] \le \sqrt{ E_{\pi}[ \|X\|^2]}<\infty$, $\|x\|$ is integrable
with respect to $\pi$ and so the absolute continuity of the integral implies
that
\begin{equation}
  \label{eqabsc1a}
\lim_{n\to \infty}     \int_{B^n_i\triangle B_i} \|x\| \, \pi(dx)   =0.
\end{equation}

To bound the first term in \eqref{eqabsc}, have for any  $L>0$
  \begin{eqnarray}
   &&  \Big\| \int_{B^n_i} x  \, \pi_n(dx) -  \int_{B^n_i}x \,  \pi(dx)  \Big\| \nonumber \\
    && \le       \Big\| \int_{\mathbb{R}^d}  x 1_{B^n_i \cap \{\|x\|\le L\}}  \, \pi_n(dx) \nonumber \\
    && \qquad \qquad -
         \int_{\mathbb{R}^d} x1_{B^n_i \cap \{\|x\|\le L\}}   \,  \pi(dx)
         \Big\|  \nonumber  \\
    & & \mbox{} +   \int_{\mathbb{R}^d} \|x\|1_{B^n_i \cap \{\|x\|> L\}}   \, \pi_n(dx) \nonumber \\
    && \qquad \qquad+
        \int_{\mathbb{R}^d} \|x\|1_{B^n_i \cap \{\|x\|> L\}}  \,  \pi(dx).    \label{eqabsc1} 
  \end{eqnarray}
Letting $Y_n=\|U_n\|$ and $Y=\|U\|$, where the $\mathbb{R}^d$-valued random variables $U_n$
and $U$ are distributed according to $\pi_n$ and $\pi$, respectively,  the
order-2 Wasserstein convergence of $\pi_n$ to $\pi$ in $\mathcal{Z}$  implies
that  $E[Y_n^2] \to E[Y]<\infty$ so that $\{Y_n\}$ is an $L_2$-bounded sequence
and therefore it is uniformly integrable \cite{Dud02}. Therefore
\[
\lim_{L\to \infty} \sup_{n\ge 1}  \Big( \int_{\mathbb{R}^d} \|x\|1_{B^n_i \cap \{\|x\|> L\}}   \, \pi_n(dx) \]
\[ \qquad \qquad +
        \int_{\mathbb{R}^d} \|x\|1_{B^n_i \cap \{\|x\|> L\}}  \,  \pi(dx)\Big) =0.
\]
Moreover, since each component of $ x1_{B^n_i \cap \{\|x\|\le L\}} \in \mathbb{R}^d$
is bounded by $L$ and since $\|\pi_n-\pi\|_{TV}\to 0$, the definition
\eqref{eqtotvar} of the total variation distance implies that for any fixed $L$, 
the first term on the right hand side of equation \eqref{eqabsc1} converges to zero as $n\to
\infty$. In summary, for any $\epsilon>0$ there is an $L>0$ such that the right hand
side of \eqref{eqabsc1} is less than $\epsilon$ for all $n$ large enough. This
and \eqref{eqabsc1a} then give that the leftmost term in \eqref{eqabsc}
converges to zero, i.e.,
\[
\lim_{n\to \infty}  \int_{B^n_i} x  \, \pi_n(dx) =  \int_{B_i}x \,  \pi(dx).
\]
Since $\pi_n \rightarrow \pi$ implies   $\pi_n(B_i^n)\to \pi(B_i)$, we
obtain the lemma statement for all $i$ such that $\pi(B_i)>0$. 
\end{proof}
	
\begin{proof}[Proof of Lemma \ref{ctsc}] To prove the first statement, assume
  that $(\pi_n,Q_n) \to (\pi,Q)$ in ${\cal Z} \times {\cal Q}_c$. Then for any
  $L > 0$
\begin{align}
&	\liminf_{n \rightarrow \infty} c(\pi_n,Q_n)= \liminf_{n \rightarrow
                                                     \infty} \int
                                                     \|x-\gamma_n(Q_n(x))\|^2\pi_n(dx)
                                                     \nonumber \\
	&\quad \geq \sum_{i=1}^{M} \liminf_{n \rightarrow \infty} \int_{B_i^n}
   \|x-\gamma_n(i)\|^2 1_{\{\|x\|^2 \leq L\}} \pi_n(dx)    \nonumber \\
	&\quad= \sum_{i=1}^{M} \int_{B_i} \|x-\gamma(i)\|^2 1_{\{\|x\|^2 \leq L\}} \pi(dx) \label{eq73}\\
	&\quad=\int_{\{\|x\|^2 \leq L\}} \|x-\gamma(Q(x))\|^2\pi(dx),   \nonumber
\end{align}
since $\gamma_n(i) \to \gamma(i)$ for any $i$ with $\pi(B_i)>0$ by
Lemma~\ref{receiverLemma} and so  $\gamma_n$  is bounded and so
$\|x-\gamma_n(i)\|^2 1_{\{\|x\|^2 \leq L\}}$ is uniformly bounded. This together
with the fact  $\pi_n \to \pi$ in total variation,
$\pi_n(B_i^n) \to \pi(B_i)$ and $\pi_n(B_i^n \Delta B_i) \to 0$, implies the
equality in (\ref{eq73}). Since $L$ is arbitrary, taking the limit as
$L \to \infty$ then we get
		\begin{equation}
		\liminf_{n \rightarrow \infty} c(\pi_n,Q_n) \geq
                c(\pi,Q). \nonumber 
		\end{equation}

  The proof of the continuity of $c(\pi,Q)$ in $Q$ is similar to \cite[Lemma
  7]{YukLinZeroDelay}. Assume $(Q_n,\pi)\to (Q,\pi)$. First observe that by
  Lemma~\ref{receiverLemma}, we have $\gamma_n(i) \to \gamma(i)$ for any
  $i\in \{1,\ldots,M\}$ such that $\pi(B_i)>0$. Let
  $I=\{i\in \{1,\ldots,M\}: \pi(B_i)>0\}$. Then we have for all $i\in I$,
\begin{align}
D_i \coloneqq \sup_{n\geq 1}\| \gamma_n(i)\| <\infty. \nonumber 
\end{align}
Letting $D=\max\limits_{i\in I} D_i$,  by the parallelogram law we have for all
  $i\in I$, 
\begin{align}
\|x-\gamma_n(Q_n(x))\|^2\leq 2\|x\|^2+2\|\gamma_n(Q_n(x))\|^2 \leq 2\|x\|^2 + 2
  D. \nonumber 
\end{align}
Since $\pi \in {\cal Z}$ has finite second moment, we obtain 
\begin{align}\label{truncatedL}
&\lim_{L \to \infty} \sup_{n\geq 1} \int_{B_i^n} \|x-\gamma_n(Q_n(x))\|^2
  1_{\{\|x\|^2 > L \}}  \pi(dx) \nonumber \\
 & \leq  \lim_{L \to \infty}  \int_{\mathbb{R}^d}
  \left(2\|x\|^2 + 2D \right) 1_{\{\|x\|^2 > L \}}  \pi(dx) = 0.
\end{align}
When $\|x\|^2\leq L$, $\|x-\gamma_n(Q_n(x))\|^2$ is uniformly bounded and since
$ \pi(B_i^n \Delta B_i) \to 0$, we have 
\begin{align}\label{boundedL}
	&	\int_{B_i^n} \|x-\gamma_n(Q_n(x))\|^2 1_{\{\|x\|^2 \leq L \}}  \pi(dx) \\
	& \quad \quad	\to \int_{B_i} \|x-\gamma(Q(x))\|^2 1_{\{\|x\|^2 \leq L \}}  \pi(dx).
\end{align}
Then, using truncation by $L$ together with (\ref{truncatedL}) and (\ref{boundedL}) we obtain 
\begin{align}
\int_{B_i^n} \|x-\gamma_n(Q_n(x))\|^2  \pi(dx) \to \int_{B_i}
  \|x-\gamma(Q(x))\|^2   \pi(dx) \nonumber 
\end{align}
and therefore $c(Q_n,\pi)\to c(Q,\pi)$ as $n\to \infty$.
\end{proof}

\begin{proof}[Proof of Lemma \ref{ctsKernel}]
Consider the conditional probability distribution given by  \cite{YukLinZeroDelay}
\begin{align*}\label{piHat}
&\hat{\pi}(i,\pi,Q)(A)\coloneqq P(x_{t+1} \in A|\pi_t=\pi,Q_t=Q,q=i) \\
&\quad =\frac{1}{\pi(B_i)}\int_{A}\left(\int_{B_i} \pi(dx)\phi(z|x)\right)dz,
\end{align*}
for $i \in I$ (see also  \cite{YukLinZeroDelay}). We have 
\begin{align*}
&\bigg|\int_{\mathcal{Z}}  g(\pi')P(d\pi'|\pi_n,Q_n) - \int_{\mathcal{Z}}  g(\pi')P(d\pi'|\pi,Q) \bigg|\\ 
&= \bigg|\sum_{i=1}^{M}\bigg(g(\hat{\pi}(i,\pi_n,Q_n))P(\hat{\pi}(i,\pi_n,Q_n)|\pi_n,Q_n) \\
& \qquad \qquad - g(\hat{\pi}(i,\pi,Q))P(\hat{\pi}(i,\pi,Q)|\pi,Q)\bigg) \bigg| \\
&= \bigg|\sum_{i=1}^{M}\bigg(g(\hat{\pi}(i,\pi_n,Q_n))\pi_n(Q_n^{-1}(i)) \\
& \qquad \qquad - g(\hat{\pi}(i,\pi,Q))\pi(Q^{-1}(i))\bigg) \bigg| \\
&= \bigg|\sum_{i=1}^{M}\bigg(g(\hat{\pi}(i,\pi_n,Q_n))\pi_n(B_i^n) - g(\hat{\pi}(i,\pi,Q))\pi(B_i)\bigg) \bigg|.
\end{align*}
Thus in view of the fact that $\pi_n(B_i^n) \rightarrow \pi(B_i)$ and that $g(
\cdot)$ is a continuous function, it is enough to prove that for all $i \in I$, $\hat{\pi}(i,\pi_n,Q_n) \rightarrow \hat{\pi}(i,\pi,Q)$. In turn, this is implied by \\
\begin{align*}
&\bigg|\int_{B_i^n} \pi_n(dx)\phi(z|x) - \int_{B_i} \pi(dx)\phi(z|x) \bigg| \\
&\leq \bigg|\int_{\mathbb{R}^d} (1_{B_i^n}-1_{B_i}) \pi_n(dx)\phi(z|x)\bigg| \\
&\quad + \left|\int_{\mathbb{R}^d} 1_{B_i}(\pi_n(x)-\pi(x))\phi(z|x)dx \right| \\
&\leq C (\pi_n(B_i^n \Delta B_i)+\|\pi_n-\pi\|_{TV}) \rightarrow 0,
\end{align*}
where $C$ is the uniform upper bound on $\phi$ and by Lemma \ref{Lem2LY2014} we
have $\|\pi_n-\pi\|_{TV} \to 0$.

The proof that $Pg(\pi,Q)$ is continuous in $Q$ for every fixed $\pi$ follows
from the proof above by setting $\pi_n = \pi$ for all $n$  and noting that  in
this case that the argument only requires the continuity of $g$ but not its
boundedness.

\end{proof}

	\bibliographystyle{ieeetr}

\begin{IEEEbiographynophoto}{\bf Meysam Ghomi} received the
  B.Sc. degree in Aerospace Engineering from Sharif University of
  Technology, Tehran, Iran, in 2015, an M.S. degree in Electrical and
  Electronics Engineering from Bilkent University, Ankara, Turkey,
  in 2018, and an M.S. degree in Mathematics and Statistics from Queen's University, Canada, in 2021. His research
  interests include stochastic control, information theory, and
  autonomous systems. He is currently working for the Canadian startup
  company Mojow, developing autonomous solutions for farming
  applications. \end{IEEEbiographynophoto}

\begin{IEEEbiographynophoto}{\bf Tam\'{a}s Linder} (S'92-M'93-SM'00-F'13) received the M.S. degree in electrical engineering from the Technical University of Budapest, Hungary, in 1988, and the Ph.D degree in electrical engineering from the Hungarian Academy of Sciences in 1992.
He was a post-doctoral researcher at the University of Hawaii in 1992 and a Visiting Fulbright Scholar at the Coordinated Science Laboratory, University of Illinois at Urbana-Champaign during 1993-1994. From 1994 to 1998 he was a faculty member in the Department of Computer Science and Information Theory at the Technical University of Budapest. From 1996 to 1998 he was also a visiting research scholar in the Department of Electrical and Computer Engineering, University of California, San Diego. In 1998 he joined Queen's University where he is now a Professor of Mathematics and Engineering in the Department of Mathematics and Statistics. His research interests include communications and information theory, source coding and vector quantization, machine learning, and statistical pattern recognition.
Dr. Linder received the Premier's Research Excellence Award of the Province of Ontario in 2002 and the Chancellor's Research Award of Queen's University in 2003. He was an Associate Editor for Source Coding of the IEEE Transactions on Information Theory in 2003-2004.\end{IEEEbiographynophoto}

\begin{IEEEbiographynophoto}{\bf Serdar Y\"uksel} (S'02, M'11) received his B.Sc. degree in Electrical and Electronics Engineering from Bilkent University; M.S. and Ph.D. degrees in Electrical and Computer Engineering from the University of Illinois at Urbana-Champaign in 2003 and 2006, respectively. He was a post-doctoral researcher at Yale University before joining the Department of Mathematics and Statistics at Queen's University. His research interests are on stochastic control, decentralized control, information theory, and probability. He has been an Associate Editor for the IEEE Transactions on Automatic Control, Automatica, Systems and Control Letters, and Mathematics of Control, Signals and Systems.\end{IEEEbiographynophoto}

\end{document}